\documentclass[%
 reprint,
 amsmath,amssymb,
 aps,
]{revtex4-2}

\usepackage{graphicx}
\usepackage{dcolumn}
\usepackage{bm}

\usepackage{makecell}
\usepackage{multirow}
\usepackage{wasysym}
\usepackage{pifont}
\usepackage{float}
\usepackage{placeins}

\usepackage{hyperref}
\hypersetup{colorlinks=true, linkcolor=blue, citecolor=blue, filecolor=magenta, urlcolor=blue}

\newcommand{\figRef}[2]{\ref{#1}\hyperref[#1]{#2}}

\newcommand{\tickMark}{\ding{51}}
\newcommand{\crossMark}{\ding{55}}

\ExplSyntaxOn
\NewDocumentCommand \colVec { m }
  {
    \begin{pmatrix}
      \clist_use:nn {#1} { \\ }
    \end{pmatrix}
  }
\ExplSyntaxOff

\newcommand{\angleaverage}[1]{\ensuremath{\left\langle #1 \right\rangle}}

\makeatletter
\newcounter{savesection}
\newcounter{apdxsection}
\renewcommand\appendix{\par
  \setcounter{savesection}{\value{section}}%
  \setcounter{section}{\value{apdxsection}}%
  \setcounter{subsection}{0}%
  \gdef\thesection{\@Alph\c@section}}
\newcommand\unappendix{\par
  \setcounter{apdxsection}{\value{section}}%
  \setcounter{section}{\value{savesection}}%
  \setcounter{subsection}{0}%
  \gdef\thesection{\@arabic\c@section}}
\makeatother

\allowdisplaybreaks

\begin{document}

\title{Designing Topological High-Order Van Hove Singularities: Twisted Bilayer Kagom\'{e}}

\author{David T. S. Perkins}
\author{Anirudh Chandrasekaran}
\author{Joseph J. Betouras}
\affiliation{Department of Physics, Loughborough University, Loughborough LE11 3TU, England, United Kingdom}


\begin{abstract}
    The interplay of high-order Van Hove singularities and topology plays a central role in determining the nature of the electronic correlations governing the phase of a system with unique signatures characterising their presence. Layered van der Waals heterostuctures are ideal systems for band engineering through the use of twisting and proximity effects. Here, we use symmetry to demonstrate how twisted Kagom\'{e} bilayers can host topological high-order Van Hove singularities. We study a commensurate system with a large twist angle and demonstrate how the initial choice of high-symmetry stacking order can greatly influence the electronic structure and topology of the system. We, furthermore, study the possibility of sublattice interference in the system. Our results illustrate the rich energy landscape of twisted Kagom\'{e} bilayers and unveil large Chern numbers (of order 10), establishing twisted bilayer Kagom\'{e} as a natural playground for probing the mixing of strong correlations and topology.
\end{abstract}

\maketitle

\section{Introduction}

Electronic instabilities are facilitated when the kinetic energy of electrons is dominated by the characteristic energy scale of their interactions, resulting in the emergence of different phases. Naturally, the study of vanishing Fermi velocities, occurrence of flat bands, and their associated divergences in the density of states (DOS) play a central role in determining the electronic phases, transport, and thermodynamic properties of materials \cite{Classen2024}. The simplest example of this is the Van Hove singularity (VHS), a saddle point within the band structure that yields a logarithmic divergence in the DOS \cite{vanHove1953}. More recently, higher-order VHSs (HOVHSs) and flat bands have attracted much attention due to their appearance in several systems exhibiting exotic phenomena: unusual thermodynamics, strange metallicity and non-Fermi liquid behavior has been observed in Sr${}_{3}$Ru${}_{2}$O${}_{7}$ \cite{Grigera2001,Tamai2008,Efremov2019,Mousatov2020} and twisted WSe${}_{2}$ bilayers \cite{Wei2024}, while unconventional superconductivity manifests in multilayer graphene both with and without twisting \cite{Cao2018,Zhou2021,Park2022,Guerci2022,Cao2021}. By tuning a system to a HOVHS, a system's phase diagram can be altered drastically, as has been shown for bernal bilayer graphene where the presence of a HOVHS enabled ferromagnetism, spin-singlet superconductivity, and charge, spin, and pair density waves \cite{Shtyk2017,Lee2024}. Every perturbation is relevant because the Fermi velocity tends to zero. HOVHSs are central in the realisation of a supermetal phase \cite{Isobe2019}. The telltale sign of a HOVHS is a more severe divergence of the DOS, scaling according to a power-law, $g(\varepsilon) \sim \varepsilon^{-\alpha}$ ($\alpha > 0$), rather than a logarithm \cite{Efremov2019,Chandrasekaran2020}, where $\varepsilon$ is the distance in energy space from the singularity. A detailed classification scheme has recently been developed for the various types of HOVHS \cite{Chandrasekaran2020,Yuan2020} in conjunction with methods to engineer and analyze their nature \cite{Chandrasekaran2022, Chandrasekaran_NatComms_2024} as well as a study of their resilience to disorder \cite{Chandrasekaran2023}. As a result, the exact connection between HOVHSs and the observed phenomena, as well as methods to engineer quantum materials with desired properties as a consequence of HOVHSs and flat bands, are areas of intense research and discussion.

An exemplary system for probing the physics of strong correlations is the Kagom\'{e} lattice, which possesses a perfectly flat band across the entire Brillouin zone (BZ), while hosting Dirac electrons around the BZ corners and VHSs at the BZ edge. The Kagom\'{e} lattice with nearest-neighbor hopping is the line graph of a honeycomb lattice. Graph theory guarantees that such lattices possess at least one flat band \cite{Lieb1989,Kollar2020,Ma2020}. At the same time, the Kagom\'{e} lattice appears in both its monolayer and bilayer forms in several materials, including the vanadium-based antimonides (AV${}_{3}$Sb${}_{5}$; A = K, Rb, Cs) \cite{Ortiz2019,Kang2022_CsVSb,Kim2023,Luo2023}, CsTi${}_{3}$Bi${}_{5}$ \cite{Li2023,Yang2024_Kagome}, and Kagom\'{e} magnets \cite{Ye2018,Yin2020,Li2021}.
In these cases, the collection of atoms forming the Kagom\'{e} lattice are embedded in a network of atoms that do not form part of the Kagom\'{e} lattice. These atomic cages prevent band engineering through methods that have proven extremely powerful in the field of van der Waals (vdW) heterostructures, where various two-dimensional (2D) materials may be stacked and offset by a relative twist angle to manifest HOVHSs and flat bands, enable strong correlations, induce topological phases, and tune proximity-induced couplings \cite{Cao2018,Park2022,Wang2020,Devakul2021,Peterfalvi2022,Veneri2022,Sun2023,Rao2023,Perkins2024,Yang2024_twist,Abouelkomsan2024}. There is a recent comprehensive study that predicts many more stoichiometric materials with Kagom\'{e} lattice structures \cite{Regnault2022}. An alternative family of 2D materials exhibiting signatures of Kagom\'{e} physics are metal-organic frameworks (MOFs), which may be constructed to host a Kagom\'{e} pattern without an atomic cage \cite{Yamada2016,Takenaka2021,Field2022,Kang2022_MOFs,Lowe2024,Fuchs2020}.

\begin{figure*}[t]
    \centering
    \includegraphics[width=\textwidth]{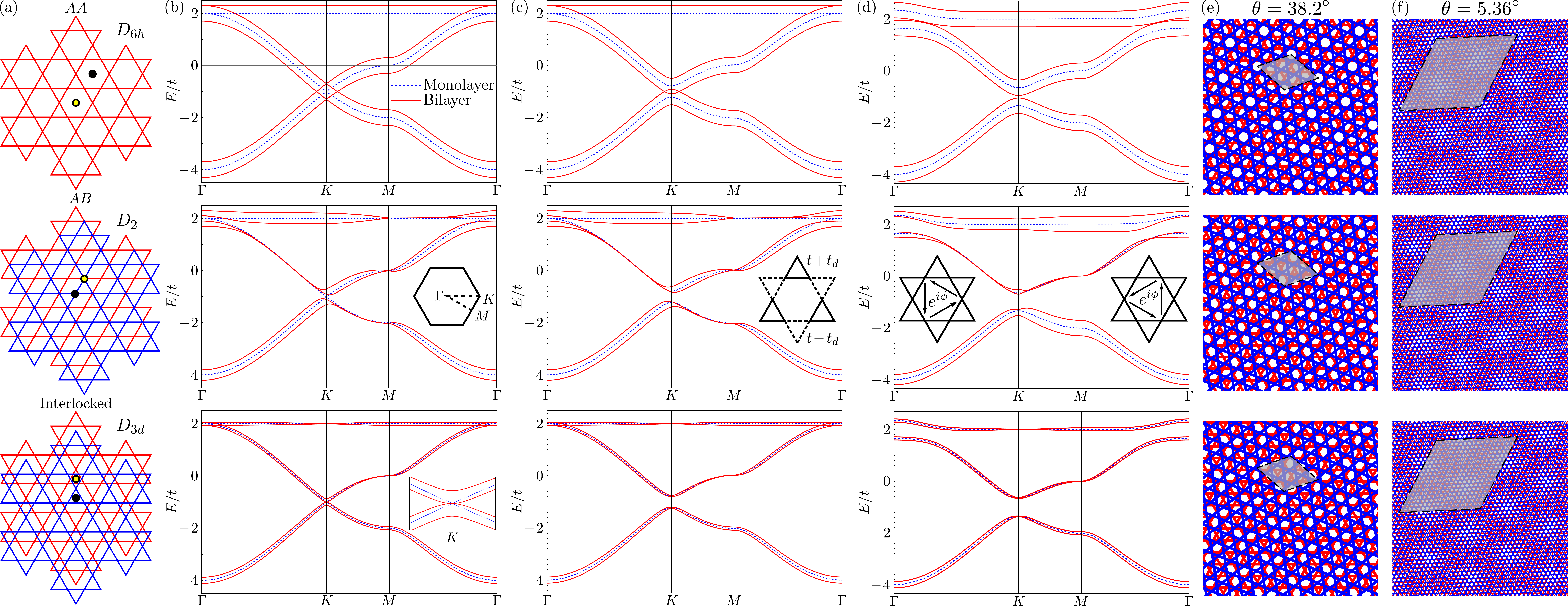}
    \caption{Unique high-symmetry stacking orders of a Kagom\'{e} bilayer and their effect upon the band structure and moir\'{e} pattern. (a): High-symmetry stacking orders with their real-space high-symmetry point (i.e. that which exhibits the full point group symmetry) highlighted by the yellow dot and twist centres used to create twisted bilayers denoted by the black dots. (b): Band structures associated to each stacking choice compared against the monolayer Kagom\'{e} bands in units of the intralayer tunneling amplitude, $t$. We use $t_{\perp} = 0.3 t$ for the interlayer tunneling energy in the $AA$ and $AB$ stackings, while taking $\tilde{t}_{\perp} = 0.2t_{\perp}$ for the reduced interlayer hopping in the interlocked scenario due to no sites overlapping directly. The inset for the interlocked case shows how the Dirac cone is flattened. (c): Band structure under dimerization ($t_{d} = 0.07 t$) for the three stackings compared to the dimerised monolayer, with the bands being gapped out around the original MDP due to breaking of inversion symmetry. The schematic in the $AB$ plot illustrates the enhancement/reduction (solid/dashed triangles) of tunneling energy due to dimerization. (d): Band structure due to complex NNN tunneling with $t_{\text{H}} = 0.1t$ and $\phi = \pi/2$ for the Haldane hopping strength and phase, respectively, where degeneracies are lifted by the breaking of $\mathcal{T}$. The schematics in the $AB$ panel shows the hopping processes yielding phase factors of $e^{i\phi}$. Reversal of hopping direction instead yields $e^{-i\phi}$. (e)-(f): Moir\'{e} patterns for TBK with $(m,n) = (5,3)$ (e) and $(m,n) = (37,3)$ (f).}
    \label{Stacking_effects}
\end{figure*}

It is therefore natural to consider Kagom\'{e}-based vdW heterostructures when designing bespoke band structures hosting HOVHSs and near-flat bands which may be further tuned through straining \cite{Consiglio2022} and twisting \cite{Lima2019}. Moreover, topological Kagom\'{e} effects will become more accessible through the plethora of potential 2D partner materials and the tunability granted by twisting, allowing for the realisation of topological HOVHSs (i.e. HOVHSs situated in bands with non-zero Chern numbers); these are expected to play a key role in the formation of phases when incorporating interactions. For example, the superconducting phase and charge ordering in AV${}_{3}$Sb${}_{5}$ are understood to stem from the topological nature of the band structure \cite{Neupert2022}, suggesting that twisting may enrich the phase space of Kagom\'{e} materials. Topological VHSs have been studied in a 3D context via Weyl metals \cite{Fontana2021}, while topological HOVHSs have only very recently been considered in a Kagom\'{e} monolayer \cite{Wang2025}.

In this work, we demonstrate how topological HOVHSs with two- ($C_{2z}$), three- ($C_{3z}$), and six-fold ($C_{6z}$) rotational symmetry can be engineered in twisted Kagom\'{e} bilayers for all possible high-symmetry stackings of the Kagom\'{e} bilayer \cite{Perkins2025}, and how the choice of stacking can greatly influence the nature and number of HOVHSs present in the band structure. We introduce the three high-symmetry stacking configurations and illustrate how the electronic structure is affected by tunneling asymmetry (dimerization) and complex next-nearest-neighbor (NNN) hopping. We then show how the structural symmetry of twisted bilayer Kagom\'{e} (TBK) is reduced upon twisting to a commensurate angle. Next, we model TBK using a tight-binding Hamiltonian that generalizes the Hamiltonian of Ref. \cite{Lima2019} by including dimerization, complex NNN tunneling, and an interlayer potential (i.e. out-of-plane electric field) to enable band engineering for creating topological HOVHSs. To establish HOVHSs and illustrate the importance of stacking order, we focus on TBK with a $38.2^{\circ}$ twist corresponding to the minimum moir\'{e} unit cell achievable.

We find that monkey saddles (third-order VHSs) appear around the moir\'{e} BZ (MBZ) corners for a large range of parameter values in several bands, while cuspoid singularities (elongated VHSs) appear around the MBZ edges without any need for tuning, and identify two classes of singularities that arise for both. Moreover, the stacking configurations with an effective $C_{6z}$ symmetry in momentum space also host very sensitive sixth-order HOVHSs at the centre of the MBZ. We use symmetry arguments to establish the exact number of singularities that may occur simultaneously within each band. Finally, when a HOVHS is created through the use of complex NNN tunneling, the breaking of time-reversal symmetry lifts the degeneracy between bands and allows us to calculate the Chern index for each one. We unveil bands with a Chern number of order 10 while simultaneously hosting a HOVHS, establishing a zoo of topological HOVHSs in TBK. The results we present here outline the electronic and topological signatures of twisted Kagom\'{e} systems and, given the growing family of Kagom\'{e} materials and the bespoke nature of vdW heterostructures, will help the design and investigation of topological Kagom\'{e} HOVHSs and nearly flat bands with exciting properties.

\section{High-Symmetry Stacking and Commensurate Twisting}

\subsection{Stacking Configurations}

An untwisted bilayer formed of two identical Kagom\'{e} lattices has three high-symmetry stacking orders (Fig. \figRef{Stacking_effects}{a}): $AA$, in which the sublattices of both layers are perfectly aligned and retains the $D_{6h}$ symmetry of the monolayer; $AB$, where one layer is shifted by a bond length relative to the other, resulting in only two directly overlapping sublattices and reducing the point group symmetry to $D_{2}$; \textit{interlocked}, reminiscent of Bernal stacked graphene, wherein the up triangles of one layer are aligned with the down triangles of the other layer to reduce the bilayer's symmetry to $D_{3d}$. The band crossing at the $K$ point of the Brillouin zone is protected by inversion symmetry, $C_{i}$, and time-reversal symmetry, $\mathcal{T}$, meaning that only the $AB$ bilayer will naturally have a gapped structure around the Dirac point of the original monolayer (Fig. \figRef{Stacking_effects}{b}). We note that the band structure around the $K$ point for the $AA$ and interlocked systems strongly resembles those of $AA$ and Bernal stacked graphene, respectively. We interpret this as a triangle exchange symmetry, like that of the sublattice exchange symmetry in graphene \cite{Mele2010}, that can be either even ($AA$) or odd (interlocked). Details of each bilayer's Hamiltonian are provided in Ref. \cite{SMref}.

To achieve monkey saddle HOVHSs around the MBZ corners, it is necessary to eliminate the linear band crossings at the Dirac points in the twisted bilayer and gain control over the emergent gaps by breaking either $C_{i}$ or $\mathcal{T}$; whether or not a Dirac cone can be flattened is ultimately determined by the symmetries of the system \cite{Sheffer2023}. Inversion symmetry is broken in breathing Kagom\'{e} lattices whose up and down triangles are of different sizes, leading to an enhanced tunneling via up triangles and a reduced tunneling via down triangles (Fig. \figRef{Stacking_effects}{c}) \cite{Guo2009,Ciola2021}. Niobium halides \cite{Regmi2023}, their chalcogen substituted counterparts \cite{Zhang2023}, and Fe${}_{3}$Sn${}_{2}$ \cite{Ye2018} are all examples of systems possessing such a lattice. To break $\mathcal{T}$, we may either apply a magnetic field or introduce a complex Haldane-type NNN tunneling \cite{Haldane1988,Wang2025}, whereby hopping anti-clockwise(clockwise) carries a phase factor of $e^{i\phi}$ ($e^{-i\phi})$ (Fig. \figRef{Stacking_effects}{d}). While both approaches will gap the Dirac cone, they ultimately yield different symmetries for the system. Dimerisation is a geometric effect and will reduce the point-group symmetry of the bilayer: $D_{6h} \rightarrow D_{3h}$, $D_{3d} \rightarrow C_{3v}$, and $D_{2} \rightarrow D_{1}$. In contrast, Haldane-type hopping possesses a $C_{6z}$ symmetry, thus not affecting the principal rotational symmetry. However, applying the in-plane dihedral rotation associated with the lattice will reverse the direction of the Haldane-type hopping ($e^{i\phi} \leftrightarrow e^{-i\phi}$). Therefore, the point-group symmetry will be reduced from $D_{6h} \rightarrow C_{6h}$, $D_{3d} \rightarrow C_{3v}$, and $D_{2} \rightarrow C_{2}$.

\subsection{Commensurate Twist Angles}

When introducing a twist into Kagom\'{e} bilayers, only certain angles will produce commensurate structures with a finite moir\'{e} unit cell (MUC), with smaller twists possessing larger MUCs and many thousands of lattice sites. Whether the bilayer is commensurate or not is determined by its underlying triangular lattice. We can label each twist angle yielding a commensurate bilayer, $\theta_{c} \in [0,60^{\circ})$, by a unique pair of coprime integers, $(m,n)$, such that \cite{Shallcross2008,Shallcross2010,Scheer2022}
\begin{equation}
    \cos \theta_{c} = \frac{3m^{2} - n^{2}}{3m^{2} + n^{2}}, \quad \sin \theta_{c} = \frac{2\sqrt{3} \, mn}{3m^{2} + n^{2}},
    \label{Commensurate_angle_definition}
\end{equation}
We apply the twist to the Kagom\'{e} bilayer by rotating one layer about a $C_{6z}$ hexagon centre to maintain $60^{\circ}$ twist periodicity, see Fig. \figRef{Stacking_effects}a, and note that this labeling of commensurate angles yields two families \cite{Scheer2022,Perkins2025}: (i) where $n$ is divisible by 3 and (ii) where $n$ is not divisible by 3. These two sets can be simply related via $\theta_{c}' = \pi/3 - \theta_{c}$ such that if $\theta_{c}$ is defined by a choice of $n$ that is divisible by 3, then $\theta_{c}'$ will be associated to a value of $n$ that is not divisible by 3 \cite{Scheer2022}. For example, $(m,n) = (5,3)$ yields $\theta_{c} = 38.2^{\circ}$ which in turn has the partner angle $\theta_{c}' = 21.8^{\circ}$ which can be obtained with $(m,n) = (3,1)$. It was shown in Ref. \cite{Perkins2025} that the point group symmetries for both families of commensurate angles were identical when the twist is applied about a hexagon centre. However, differences between the two can be seen both at the superlattice level and in the Hamiltonian.

Working in the twist-symmetric frame (i.e. layer 1 is rotated by $-\theta_{c}/2$ and layer 2 is rotated by $\theta_{c}/2$), the cases where $n$ is divisible by 3 will always possess a moir\'{e} lattice vector parallel to the $x$-axis, while those with $n$ not divisible by 3 will always possess a moir\'{e} lattice vector parallel to the $y$-axis. Using graphene nomenclature, we refer to these orientations of the unit cell as zig-zag and armchair, respectively, based upon the Wigner-Seitz cell constructed with the twist origin at its centre. To see this, let us denote the lattice sites of a triangular lattice by $\mathbf{r}_{i}(k,l) = k \mathbf{a}_{1} + l \mathbf{a}_{2}$, where $i$ is the layer index and $k,l \in \mathbb{Z}$, and their rotated forms by $\mathbf{r}_{i}^{\theta} = R_{\theta} \mathbf{r}_{i}$. In the twist-symmetric frame, the triangular lattice is guaranteed to have dihedral symmetry along the $x$-axis which manifests as a $y \rightarrow -y$ reflection symmetry between the layers. Consequently, a moir\'{e} lattice vector must lie either parallel to the $x$-axis or $y$-axis.

Without loss of generality \cite{Perkins2025}, we assume that $n$ is divisible by 3 such that $n = 3\nu$ with $\nu \in \mathbb{Z}$ and demand that the $y$-components of $\mathbf{r}_{1}^{\frac{\bar{\theta}_{c}}{2}}(k,l)$ and $\mathbf{r}_{2}^{\frac{\theta_{c}}{2}}(p,q)$ to vanish ($\bar{\theta}_{c} = -\theta_{c}$). We find that $\mathbf{r}_{1}^{\frac{\bar{\theta}_{c}}{2}}(k,l) = \mathbf{r}_{2}^{\frac{\theta_{c}}{2}}(p,q)$ is then satisfied by
\begin{equation}
\begin{gathered}
    k = \alpha_{m\nu} (m - \nu), \quad l = 2\alpha_{m\nu}\nu,
    \\
    p = \alpha_{m\nu}(m + \nu), \quad q = -2\alpha_{m\nu}\nu,
\end{gathered}
\end{equation}
where
\begin{equation}
    \alpha_{m\nu} = \begin{cases}
        1, &m + \nu \text{ is odd} \\
        \frac{1}{2},  &m + \nu \text{ is even}
    \end{cases}.
\end{equation}
The resulting moir\'{e} lattice constant is then $a_{\text{M}} = \alpha_{m\nu} \sqrt{m^{2}+3\nu^{2}} a$, where $a$ is the monolayer lattice constant. If we instead demand that the $x$-component vanish, we find a larger magnitude for the moir\'{e} lattice constant, $\tilde{a}_{\text{M}} = \alpha_{m\nu} \sqrt{3} \sqrt{m^{2}+3\nu^{2}} a$, clearly indicating that the commensurate superlattices defined by an $n$ divisible by 3 will always be of zig-zag orientation. Alternatively, we may assume that $n = 3\nu \pm 1$ and demand that the $x$-components of $\mathbf{r}_{1}^{\frac{\bar{\theta}_{c}}{2}}(k,l)$ and $\mathbf{r}_{2}^{\frac{\theta_{c}}{2}}(p,q)$ to vanish, where we find that $\mathbf{r}_{1}^{\frac{\bar{\theta}_{c}}{2}}(k,l) = \mathbf{r}_{2}^{\frac{\theta_{c}}{2}}(p,q)$ is satisfied when
\begin{equation}
\begin{gathered}
    k = \alpha_{mn} (m + n), \qquad l = -2\alpha_{mn} m,
    \\
    p = \alpha_{mn} (m - n), \qquad q = -2\alpha_{mn} m,
\end{gathered}
\end{equation}
yielding a moir\'{e} lattice constant of $a_{\text{M}} = \alpha_{mn} \sqrt{3 m^{2}+n^{2}} a$. Similar to the $n = 3\nu$ case, we find a larger moir\'{e} lattice constant, $\tilde{a}_{\text{M}} = \alpha_{mn} \sqrt{3} \sqrt{3 m^{2}+n^{2}} a$, by seeking a moir\'{e} lattice vector parallel to the $x$-axis when $n$ is not divisible by 3. Therefore, the orientation of the superlattice unit cell when 3 does not divide $n$ will always be armchair. Lastly, we note that these two classes of superlattices are related by reflection in the $y = x$ line, see Appendix \ref{Moire_class_sec} for details.

Moving onto the Hamiltonians of these systems, in the absence of complex NNN hopping and dimerization, we see that the Hamiltonian with $\theta_{c}$, $\mathcal{H}$, is related to the Hamiltonian with $\theta_{c}'$ by $\mathcal{H}(k_{x},k_{y}) = \mathcal{H}'(k_{y},k_{x})$ when constructed in the twist-symmetric frame. Application of an out-of-plane electric field, which induces an interlayer potential, does not affect this observation. Including Haldane-type hopping, we must also flip the sign of the phase to obtain one Hamiltonian from the other, $\mathcal{H}(k_{x},k_{y},\phi) = \mathcal{H}'(k_{y},k_{x},-\phi)$. Finally, dimerization results in the largest difference between the two systems, with the asymmetric tunneling along up and down triangles appearing reversed in one layer and unchanged in the other after reflection in $y = x$. Specifically, the layer rotated anti-clockwise by $\theta_{c}/2$ will appear to have its dimerization unchanged while the layer rotated clockwise by $\theta_{c}/2$ will have its dimerization reversed, details are discussed in Appendix \ref{Moire_class_sec}. Therefore, the two families of twist angles may provide significant differences when dimerization is present.

Finally, while the choice of stacking order might initially seem irrelevant for small twist angles, it becomes apparent at large twist angles that the superlattice structure is completely dependent on the stacking order (Fig. \figRef{Stacking_effects}{e,f}). The point group symmetry of the resulting superlattice will always be dihedral and retain the rotational symmetry of the original untwisted Kagom\'{e} bilayer: $D_{6}$ for AA, $D_{3}$ for interlocked, and $D_{2}$ for AB \cite{Perkins2025}. This is further reflected in the electronic structure, where parabolic band touching at the $K$ point is seen for interlocked TBK while a double well with a linear crossing is seen for $AA$ TBK.

\begin{figure*}[t]
    \centering
    \includegraphics[width=\textwidth]{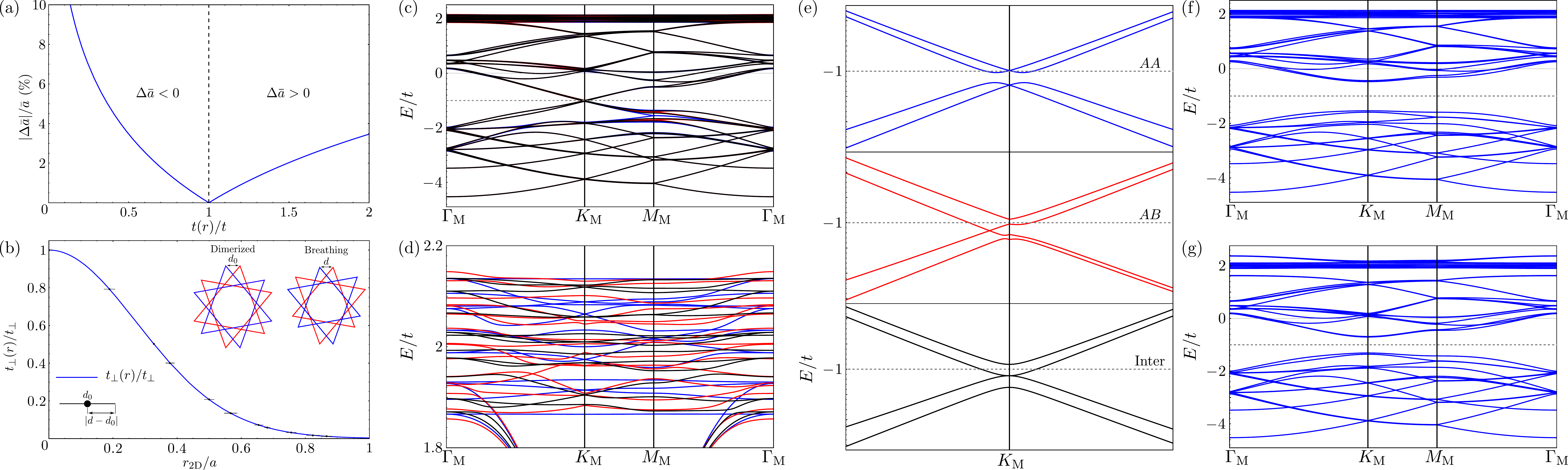}
    \caption{Variation of the TBK band structure due to stacking, dimerization, and Haldane-type hopping. Here we consider a TBK system with $\theta_{c} = 38.2^{\circ}$, $a = 0.5338$ nm, $d_{\perp} = 0.6596$ nm, $t_{\perp} = 0.3 t$, and $\gamma = 20$ \cite{Ye2018,Lima2019}. (a): Change in NN separation distance required to achieve a given decay factor for intralayer tunneling. (b): Plot of the interlayer tunnelling decay as a function of the in-plane separation of atoms, $r_{\text{2D}}$. The black dots correspond to the separations present within a twisted dimerized Kagom\'{e} bilayer with $\theta_{c} \simeq 38.2^{\circ}$, and the error bars indicate the magnitude of difference in 2D interlayer separations between the dimerized and breathing Kagom\'{e} bilayers such that they both yield a difference in tunnelling along up and down triangles of $t_{d} = 0.5 t$. (c)-(e): Band structure in the absence of dimerization and Haldane hopping for all high-symmetry stackings (blue: $AA$, red: $AB$, black: interlocked), with a focus on the near-flat band region (d) and on the MDP energy (e) marked by the grey dashed line. (f): The effect of dimersiation on $AA$ TBK with $t_{d} = 0.2 t$, opening a gap between bands 14 and 15. Similar band structures for $AB$ and interlocked TBK are obtained, with differences in stacking order becoming more apparent over a wider range of the high-symmetry path, see Ref. \cite{SMref}. (g): $AA$ TBK with $t_{\text{H}} = 0.1 t$ which opens two gaps: one between bands 14 and 15, and a second between bands 28 and 29.}
    \label{Twisting_and_tunneling}
\end{figure*}

\section{Model}

\subsection{Tight-Binding Hamiltonian}

We model the TBK system using a tight-binding model with one atomic orbital per lattice site, assuming an exponential decay of the tunneling energy with separation distance, $r$, of the form $f(r) = \exp(-\gamma(r-r_{i})/r_{i})$, where $r_{i}$ is the NN separation ($\bar{a} = a/2$) for intralayer tunneling and the interlayer separation ($d_{\perp}$) for interlayer tunneling, while $\gamma$ characterises the hopping range \cite{Lima2019}. By retaining tunneling between all lattice sites up to NNN MUCs, the resulting momentum space Hamiltonian becomes:
\begin{equation}
\begin{split}
    H = &- \sum_{\mathbf{k}} \sum_{l} \sum_{\alpha,\beta} \left[ t^{\alpha\beta} \Gamma_{l}^{\alpha\beta}(\mathbf{k}) - s_{l}^{\null} \frac{\Delta}{2} \delta_{\alpha\beta} \right] c_{\mathbf{k}l\alpha}^{\dagger} c_{\mathbf{k}l\beta}^{\null}
    \\
    &- \sum_{\mathbf{k}} \sum_{\alpha,\beta} t_{\perp}^{\alpha\beta} \left[ \Gamma_{\perp}^{\alpha\beta}(\mathbf{k}) c_{\mathbf{k}1\alpha}^{\dagger} c_{\mathbf{k}2\beta}^{\null} + \text{h.c.} \right],
\end{split}
\end{equation}
where $c_{\mathbf{k}l\alpha}^{\dagger}$ ($c_{\mathbf{k}l\alpha}^{\null}$) is the creation (annihilation) operator for an electron with momentum $\mathbf{k}$ in layer $l$ for moir\'{e} sublattice site $\alpha$, $t^{\alpha\beta} = t$ is the intralayer tunneling amplitude between sites $\alpha$ and $\beta$, $t_{\perp}^{\alpha\beta} = t_{\perp}^{\null}$ is the interlayer tunneling amplitude between sites $\alpha$ and $\beta$ of different layers, $\Delta$ is an interlayer potential, $s_{l} = \pm 1$ for layer 1/2, $\delta_{\alpha\beta}$ is the Kronecker delta, and $\Gamma_{l(\perp)}^{\alpha\beta}(\mathbf{k})$ are the structure factors containing the information about the decay of tunneling energy with distance (their explicit form is given in Appendix \ref{Hamiltonian_derivation_sec}). To include dimerization, we restrict intralayer hopping to only NN and let $t^{\alpha\beta} = t \pm t_{d}$ with the sign determined by whether the tunneling is via an up or down triangle. For the models we consider here with $\gamma = 20$, NN intralayer tunneling is sufficient to accurately model the system, since NNN intralayer hopping introduces negligible corrections $\sim 10^{-7} t$. Additionally, we may introduce an intralayer Haldane-type hopping via
\begin{equation}
    H_{\text{H}}^{\null} = - t_{\text{H}}^{\null} \sum_{\mathbf{k}} \sum_{l} \sum_{\alpha,\beta} \Gamma_{\text{H}}^{\alpha\beta}(\mathbf{k}) e^{i \mathcal{S}_{\alpha\beta}\phi} c_{\mathbf{k}l\alpha}^{\dagger}c_{\mathbf{k}l\beta}^{\null},
    \label{Haldane_term}
\end{equation}
where $\mathcal{S}_{\alpha\beta} = \pm 1$ and the details of the Haldane structure factor, $\Gamma_{\text{H}}^{\alpha\beta}(\mathbf{k})$, are given in Appendix \ref{Hamiltonian_derivation_sec}. Finally, we note that the Haldane hopping will not be subject to dimerization due to being a NNN tunneling process. The total Hamiltonian is thus given by $\mathcal{H}(t_{d},t_{\text{H}},\phi,\Delta;\theta) = H + H_{\text{H}}$. A detailed derivation of the total Hamiltonian is given in Appendix \ref{Hamiltonian_derivation_sec}.

\subsection{Electronic Structure and Stacking}

We illustrate the effects of stacking order, dimerization, and Haldane-type hopping in Fig. \ref{Twisting_and_tunneling} for a TBK system based on Fe${}_{3}$Sn${}_{2}$ \cite{Ye2018,Lima2019}. For a physical interpretation of dimerization in TBK systems, we note that dimerizations as large as $t_{d} \sim 0.5 t$ are achievable with small changes to the interatomic distances on the order of a 2-3\% (Fig. \figRef{Twisting_and_tunneling}{a}). Therefore, the dimerization term approximates well a TBK system comprised of breathing Kagom\'{e} lattices at large twist angles. The effective separation error compared to the modelled interlayer tunneling decay is presented in Fig. \figRef{Twisting_and_tunneling}{b}. Small changes to the tunneling energies will only renormalize the values of the parameters needed to tune a HOVHS.

Starting from $t_{d} = t_{\text{H}} = 0$, we see that the stacking order is particularly relevant around the $M_{\text{M}}$ point with clear differences in both energy and structure (Fig. \figRef{Twisting_and_tunneling}{c}). We further observe the appearance of a characteristic densely packed near-flat band region about the original monolayer flat band energy with a resolution of $\sim 0.3 t$, where the stacking order greatly influences the fine details of each band, although no choice of stacking order prevents the loss of an exact flat band (Fig. \figRef{Twisting_and_tunneling}{d}). The nature of the electronic structure of the aligned bilayer at the $K_{\text{M}}$ point is also preserved upon twisting to a commensurate angle (Fig. \figRef{Twisting_and_tunneling}{e}). Specifically, the gap and linear band crossing of the $AA$ system persists while the interlocked stacking retains its parabolic band touching at the $K_{\text{M}}$ point, with both sets of features occurring close to the original monolayer Dirac point (MDP) energy. These features prevent us from observing critical points that also possess a vanishing curvature in the absence of dimerization or Haldane-type hopping.

\begin{table*}
\centering
\caption{A summary of the maximum number of simultaneous HOVHSs allowed around the high-symmetry points of the MBZ for $AA$ and interlocked TBK. The entries in the left section of the table indicate which tuning parameters are present, while the middle section presents the magnetic point group for each case using international (left) and Sch\"{o}nflies (right) notation. We indicate the axes of dihedral symmetry explicitly in the superscripts of the crystallographic dihedral point groups when working in the twist-symmetric frame. The right section of the table lists the number of two- (second-order and fourth-order), three- (monkey saddle), and six-fold (sixth-order) singularities permitted to occur simultaneously at the MBZ high-symmetry points for that combination of parameters for the two stacking orders. The subscripts denote the MUC orientation ($\hexagon$ for armchair and $\varhexagon$ for zig-zag) and stacking order where their number of singularities differ.}
\label{Symmetry_table}
\begin{tabular}{ c | c | c || c  c | c  c | c  c || c }
        $t_{d}$ & $t_{\text{H}}$ & $\Delta$ & \multicolumn{2}{c|}{Zig-zag/Armchair $AA$} & \multicolumn{2}{c|}{Zig-zag Interlocked} & \multicolumn{2}{c||}{Armchair Interlocked} & HOVHSs ($C_{2z}$, $C_{3z}$, $C_{6z}$) \\
    \hline\hline
        \crossMark & \crossMark & \crossMark & $6 \, 2 \, 2$ & $D_{6}^{(x,y)} \cup \mathcal{T}$ & $3 \, 2 \, 1$ & $D_{3}^{(x)} \cup \mathcal{T}$ & $3 \, 1 \, 2$ & $D_{3}^{(y)} \cup \mathcal{T}$ & (3,2,1) \\
    \hline
        \tickMark & \crossMark & \crossMark & $3 \, 1 \, 2$ & $D_{3}^{(y)} \cup \mathcal{T}$ & $3$ & $C_{3}^{\null} \cup \mathcal{T}$ & $3 \, 1 \, 2$ & $D_{3}^{(y)} \cup \mathcal{T}$ & (3,2,1) \\
    \hline
        \crossMark & \tickMark & \crossMark & $6 \, 2' \, 2'$ & $(1 + \mathcal{T} C_{2x}^{\null}) C_{6}^{\null}$ & $3 \, 2' \, 1$ & $(1 + \mathcal{T} C_{2x}^{\null}) C_{3}^{\null}$ & $3 \, 1 \, 2'$ & $(1 + \mathcal{T}C_{2y}^{\null}) C_{3}^{\null}$ & (3,2,1) \\
    \hline
        \tickMark & \tickMark & \crossMark & $3 \, 1 \, 2'$ & $(1 + \mathcal{T}C_{2y}^{\null})C_{3}^{\null}$ & $3$ & $C_{3}^{\null}$ & $3 \, 1 \, 2'$ & $(1 + \mathcal{T}C_{2y}^{\null}) C_{3}^{\null}$ & $(3,2,1)_{\hexagon}$, $(3,1,0)_{\varhexagon}$ \\
    \hline
        \crossMark & \crossMark & \tickMark & $6$ & $C_{6}^{\null} \cup \mathcal{T}$ & $3$ & $C_{3}^{\null} \cup \mathcal{T}$ & $3$ & $C_{3}^{\null} \cup \mathcal{T}$ & (3,2,1) \\
    \hline
        \tickMark & \crossMark & \tickMark & $3$ & $C_{3}^{\null} \cup \mathcal{T}$ & $3$ & $C_{3}^{\null} \cup \mathcal{T}$ & $3$ & $C_{3}^{\null} \cup \mathcal{T}$ & (3,2,1) \\
    \hline
        \crossMark & \tickMark & \tickMark & $6$ & $C_{6}^{\null}$ & $3$ & $C_{3}^{\null}$ & $3$ & $C_{3}^{\null}$ & $(3,2,1)_{AA}$, $(3,1,0)_{\text{Int}}$ \\
    \hline
        \tickMark & \tickMark & \tickMark & $3$ & $C_{3}^{\null}$ & $3$ & $C_{3}^{\null}$ & $3$ & $C_{3}^{\null}$ & (3,1,0) \\
    \hline
\end{tabular}
\end{table*}

\subsection{Emergent Symmetries that enable HOVHSs}

Through dimerization and Haldane-type hopping, we can tune the gaps between previously degenerate bands at the $K_{\text{M}}$ point to engineer a monkey saddle singularity for both the $AA$ and interlocked TBK systems (Figs. \figRef{Twisting_and_tunneling}{f,g}). This type of HOVHS requires $C_{3z}$ symmetry and therefore cannot be hosted by the $AB$ configuration, and hence we shall only focus on the $AA$ and interlocked systems from hereon. Moreover, we can explore the possibility of engineering a HOVHS with six-fold rotational symmetry at $\Gamma_{\text{M}}$ in both the $AA$ and interlocked TBK systems. Naturally, this will be achieved most easily using an $AA$ stacking due to its inherent $C_{6z}$ symmetry. When $t_{d} \neq 0$ and $t_{\text{H}} = 0$ the system possesses an effective $C_{6z}$ symmetry in momentum space due to the system possessing $C_{3z}$ and $\mathcal{T}$ (the same is true for interlocked TBK). For the opposite case with $t_{d} = 0$ and $t_{\text{H}} \neq 0$, the Haldane hopping leaves the structural $C_{6z}$ symmetry intact. If both $t_{d}, t_{\text{H}} \neq 0$, then an effective $C_{6z}$ symmetry cannot emerge in either $AA$ or interlocked TBK through the combination of $C_{3z}$ and $\mathcal{T}$. Nonetheless, due to deeper underlying symmetries present in the system, this does not prevent us from creating a six-fold singularity.

Both $AA$ and interlocked TBK possess a $C_{3}$ point group symmetry when dimerization and complex NNN tunneling are included. By working in the twist-symmetric frame, the underlying moir\'{e} superlattice (i.e. neglecting dimerization and Haldane-type hopping) will always exhibit a dihedral symmetry along the $x$-axis, $y$-axis, or sometimes both. For $AA$ TBK the underlying superlattice possesses dihedral symmetry along both axes for all commensurate twist angles: $C_{2x}$ along the $x$-axis and $C_{2y}$ along the $y$-axis. In contrast, the underlying superlattice of interlocked TBK will possess only $C_{2x}$ ($C_{2y}$) for a zig-zag (armchair) orientation. These symmetries produce the following mappings $C_{2x}: (y,z;\phi) \rightarrow (-y,-z;-\phi)$ and $C_{2y}: (x,z;-\phi) \rightarrow (-x,-z;-\phi)$, which may be interpreted as $(k_{y},\phi) \rightarrow (-k_{y},-\phi)$ and $(k_{x},\phi) \rightarrow (-k_{x},-\phi)$, respectively, and an exchange of layers. Dimerisation acts as a $C_{2x}$ breaking mechanism, due to mapping of up triangles to down triangles upon a layer exchange. The topologically trivial part of the Hamiltonian with $\Delta = 0$, $H_{0}$, is therefore symmetric under time-reversal in all cases and $C_{2y}$ for $AA$ and armchair interlocked TBK when dimerization is present,
\begin{equation}
\begin{gathered}
    H_{0} = \mathcal{T} H_{0} \mathcal{T}^{-1}, \quad H_{0}^{\text{AA}} = C_{2y}^{\null} H_{0}^{\text{AA}} C_{2y}^{-1},
    \\
    H_{0}^{\text{Int},\hexagon} = C_{2y}^{\null} H_{0}^{\text{Int},\hexagon} C_{2y}^{-1},
\end{gathered} \label{Trivial_topology_ysymmetry}
\end{equation}
where the superscript denotes the stacking configuration and MUC orientation ($\hexagon$: armchair, $\varhexagon$: zig-zag). If dimerization is absent, $H_{0} \rightarrow \bar{H}_{0}$, then up- and down-triangless are equivalent, and hence the topologically trivial piece of the Hamiltonian for zig-zag interlocked TBK, retains its dihedral symmetry:
\begin{equation}
    \bar{H}_{0}^{\text{AA}} = C_{2x}^{\null} \bar{H}_{0}^{\text{AA}} C_{2x}^{-1} \qquad \bar{H}_{0}^{\text{Int},\varhexagon} = C_{2x}^{\null} \bar{H}_{0}^{\text{Int},\varhexagon} C_{2x}^{-1}.
    \label{Trivial_topology_xsymmetry}
\end{equation}
The topological contribution to the Hamiltonian is not symmetric under any diherdral rotation or time-reversal ($C_{2}' \in \{C_{2x}^{\null},C_{2y}^{\null}\}$),
\begin{equation}
    \mathcal{T} H_{\text{H}} \mathcal{T}^{-1} = H_{\text{H}}|_{\phi\rightarrow-\phi}, \quad C_{2}' H_{\text{H}} C_{2}^{\prime-1} = H_{\text{H}}|_{\phi\rightarrow-\phi},
\end{equation}
but it is symmetric under their combined transformation,
\begin{equation}
    H_{\text{H}} = \mathcal{T} C_{2}' H_{\text{H}} (\mathcal{T} C_{2}')^{-1}.
    \label{Topology_symmetry}
\end{equation}
This results in the total Hamiltonian exhibiting emergent dihedral time-reversal symmetries for $AA$ TBK and interlocked TBK (eqs. \ref{Trivial_topology_ysymmetry}, \ref{Trivial_topology_xsymmetry}, and \ref{Topology_symmetry}), $\mathcal{H}_{0} = \mathcal{T} C_{2}' \mathcal{H}_{0} (\mathcal{T} C_{2}')^{-1}$.

The combinations of $\mathcal{T}C_{2x}$ and $\mathcal{T}C_{2y}$ manifest as $k_{x} \rightarrow -k_{x}$ and $k_{y} \rightarrow -k_{y}$ symmetries, respectively, in the 2D MBZ. By expanding the energy of a given band about the origin, the $\mathcal{T} C_{2}'$ and $C_{3z}$ symmetries of the Hamiltonian restrict the expansion to
\begin{equation}
\begin{split}
    E \simeq \tilde{a}_{2} k^{2} &+ \tilde{a}_{4} k^{4} + \tilde{a}_{6} k^{6} + a_{3} k^{3} \cos(3\varphi) \\
    &+ b_{3} k^{3} \sin(3\varphi) + a_{6} k^{6} \cos(6\varphi) + \mathcal{O}(k^{8}),
    \label{Six_fold_expansion}
\end{split}
\end{equation}
where $\varphi$ is the azimuthal angle and $\tilde{a}_{n}$, $a_{n}$, and $b_{n}$ are coefficients that are functions of the tuning parameters. Therefore, symmetry allows for the emergence of a six-fold HOVHS provided that $\mathcal{T} C_{2}'$ (an effective reflection symmetry in either the $x$- or $y$-axis) and $C_{3z}$ symmetry is preserved.

\begin{figure*}[t]
    \centering
    \includegraphics[width=\textwidth]{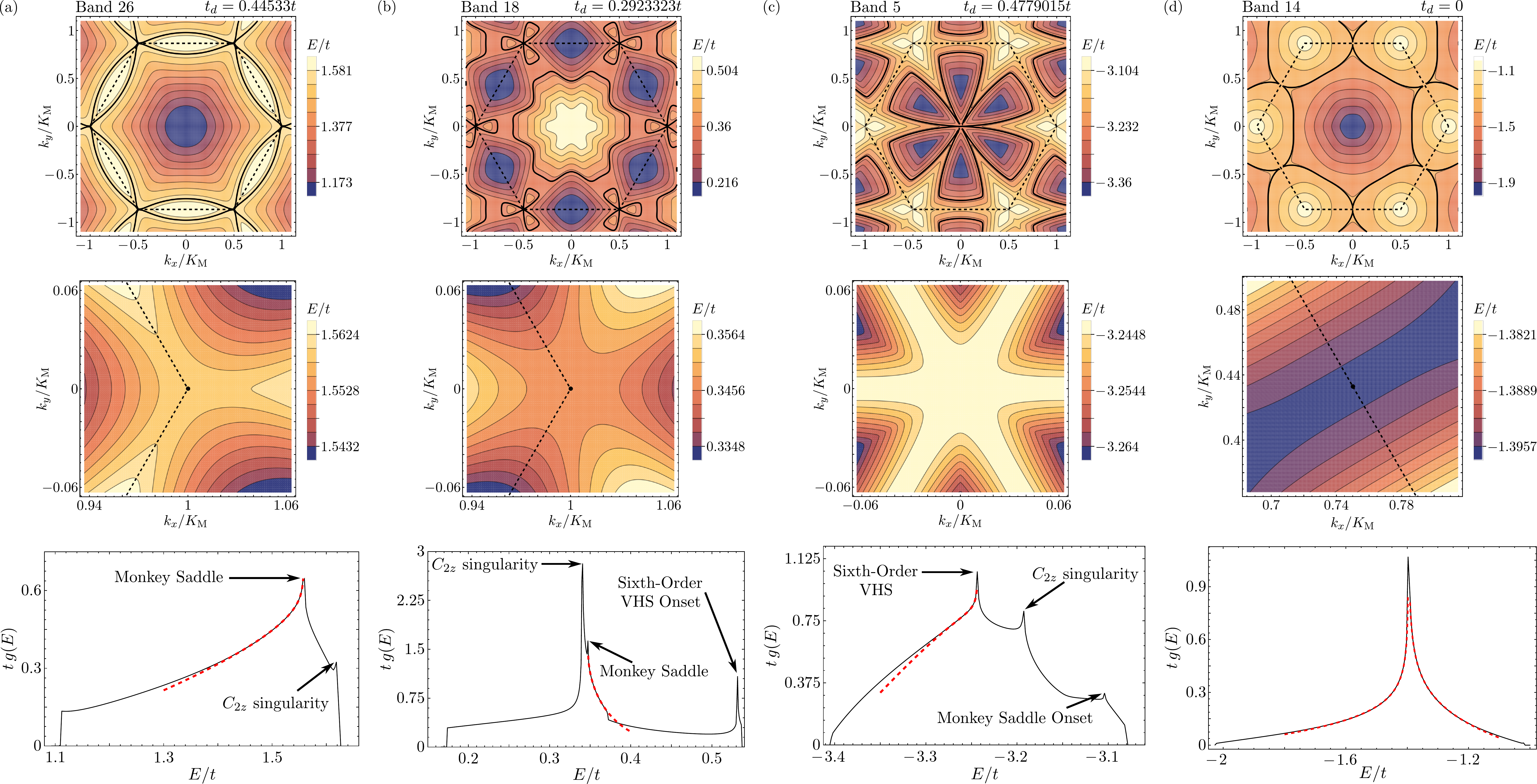}
    \caption{Energy contours of bands tuned to host HOVHSs around the MBZ high-symmetry points and their corresponding DOS. Here we present the contours of bands hosting HOVHSs at $K_{\text{M}}$ in (a) and (b), $\Gamma_{\text{M}}$ in (c), and $M_{\text{M}}$ in (d) due to dimerization, with the thick contour indicating the that critical contour of the singularity. The momentum is scaled to the $K_{\text{M}}$ point and the black dashed line indicates the boundary of the first MBZ. The top row shows the energy surfaces over the entire MBZ, while the second row focuses on the relevant HOVHS. The bottom row shows the corresponding single-spin DOS per unit volume for the band. The red dashed lines are fittings of the expected HOVHS behavior in the vicinity of the divergence (more details are provided in Ref. \cite{SMref}). The $AA$ TBK system considered here takes $\theta_{c} = 38.2^{\circ}$, $a = 0.5338$ nm, $d_{\perp} = 0.6596$ nm, $t_{\text{H}} = 0$, and $\gamma = 20$ \cite{Ye2018,Lima2019}, with $t_{\perp} = 0.3 t$ in (a), (b), and (c) and $t_{\perp} = 0.267 t$ in (d). The band number and choice of $t_{d}$ are given at the top of each column.}
    \label{HOVHS_examples}
\end{figure*}

Taking this symmetry analysis one step further, we determine how many singularities exist at the same energy. In the twist-symmetric frame, we note that the zig-zag and armchair orientations of the MUC may yield different numbers of HOVHSs possessing the same energy due to the difference in their dihedral symmetries. Let us start by assuming that a monkey saddle exists at the $K_{\text{M}}^{\null}$ point of the MBZ. If $\mathcal{T}$ is preserved, $K_{\text{M}}^{\null}$ trivially maps onto $K_{\text{M}}'$, thereby yielding two simultaneous singularities. Introducing Haldane-type hopping breaks $\mathcal{T}$ and $C_{2}'$ symmetries, but leaves the $\mathcal{T} C_{2}'$ symmetries in tact, allowing us to again connect the differing MBZ corners for zig-zag (armchair) $AA$ and interlocked TBK using $\mathcal{T} C_{2x}$ ($\mathcal{T} C_{2y}$). Only with both Haldane-type hopping and dimerization do the two orientations differ for both stacking orders, wherein $\mathcal{T} C_{2x}$ is broken in zig-zag structures and $\mathcal{T} C_{2y}$ is preserved in armchair structures. Hence with both dimerization and complex NNN hopping, zig-zag TBK will possess only a single HOVHS at a given MBZ corner energy, while armchair TBK will still exhibit two simultaneous singularities. Moving on to the case of two-fold VHSs at the MBZ edges, the number of simultaneous singularities will always be three in $AA$ and interlocked TBK due to possessing a $C_{3z}$ symmetry. Only in $AB$ TBK will the number of simultaneous singularities differ to become two instead of three due to its lack of $C_{3z}$ symmetry. When dimerization is introduced, $AB$ TBK will always exhibit a dihedral symmetry along one of three lines: the $y$-axis (i.e. the $90^{\circ}$ line) or the $\pm 30^{\circ}$ lines. These will always act to map two differing MBZ edges onto one another while leaving one to map onto iteslf. We summarise the maximum number of singularities occurring simultaneously for each possible combination of tuning parameters in Table \ref{Symmetry_table} for $AA$ and interlocked TBK.

\section{Creating Higher-Order Van Hove Singularities}

\subsection{Monkey Saddle Singularities}

Using a wide range of dimerizations, $0.05 \lesssim t_{d}/t \lesssim 0.7$, the linear band crossings at the MBZ corners are gapped out to create monkey saddles. This results in two types of singularities. Band 26 of dimerised $AA$ TBK in Fig. \figRef{HOVHS_examples}{a} illustrates a \textit{delocalized} monkey saddle characterised by energy contours connecting the singularities at $K_{\text{M}}^{\null}$ and $K_{\text{M}}'$. In contrast, band 18 (Fig. \figRef{HOVHS_examples}{b}) hosts \textit{localized} HOVHS at the MBZ corners that are disconnected from each other. Both cases yield a divergence in the DOS that is well described by a $g(\varepsilon) \sim \varepsilon^{-1/3}$ power-law at leading order (bottom panels of Figs. \figRef{HOVHS_examples}{a,b}), where $\varepsilon$ is the energy measured from the singularity. The fits shown here account for sub-leading order corrections due to the HOVHS existing within a more complex energy landscape. The forms of these sub-leading corrections are discussed in Appendix \ref{Monkey_saddle_details_sec}. Further examples of monkey saddles in other bands for $AA$ and interlocked TBK can be found in Ref. \cite{SMref}.

Both types of singularities are of particular relevance to metals exhibiting non-Fermi liquid behavior in their transport coefficients and thermodynamic properties. The presence of remote regions where the Fermi velocity vanishes (i.e. localized singularities) has been shown to alter the electrical resistivity of materials by acting as reservoirs for current-carrying electrons to scatter to and from \cite{Mousatov2020} and heat capacity \cite{Efremov2019}. Similarly, the appearance of singularities on a single Fermi surface that has both dispersive and flat regions (i.e. delocalized singularities) have been shown to dominate the interaction-induced electron self-energy, thus governing the electronic contribution to thermal transport \cite{Stangier2022}.

\subsection{Six-Fold Singularities}

Due to its high-order nature, a six-fold singularity is very sensitive to the tuning parameters and may require several to achieve an exact HOVHS. However, symmetry can reduce the number of tuning parameters necessary to yield an effective sixth-order singularity from the perspective of experiment. Fig. \figRef{HOVHS_examples}{c} presents an example of a six-fold singularity achieved using only dimerization. Our analysis of the expansion coefficients in eq. \ref{Six_fold_expansion} reveals that band 5 has a vanishing second-order coefficient ($\tilde{c}_{2} = 0$) and a fourth-order coefficent we infer to be small in comparison to the sixth-order term, $\tilde{c}_{4} \ll c_{6}$: we obtain $\tilde{c}_{4}$ to be non-zero but observe an extremely small minimum in the band $(E(\mathbf{0}) - E(0.01\mathbf{K}_{\text{M}}))/E(\mathbf{0}) \sim 10^{-7}$. Moreover, the divergence of the DOS at the HOVHS energy is easily fitted by the $g(\varepsilon) \sim \varepsilon^{-2/3}$ scaling at leading order characteristic of a sixth-order singularity \cite{Shtyk2017,Chandrasekaran2020}, see Fig. \figRef{HOVHS_examples}{c}. The onset of a six-fold singularity can also be seen in other bands. The third peak in the DOS of band 18 (Fig. \figRef{HOVHS_examples}{b}) can be attributed to a set of six closely packed two-fold singularities: closer inspection reveals a small minimum at the origin. We provide additional examples of effective six-fold HOVHSs in Ref. \cite{SMref} for $AA$ and interlocked TBK, with the latter being tuned via an interlayer potential.

\begin{figure}[t]
    \centering
    \includegraphics[width=\linewidth]{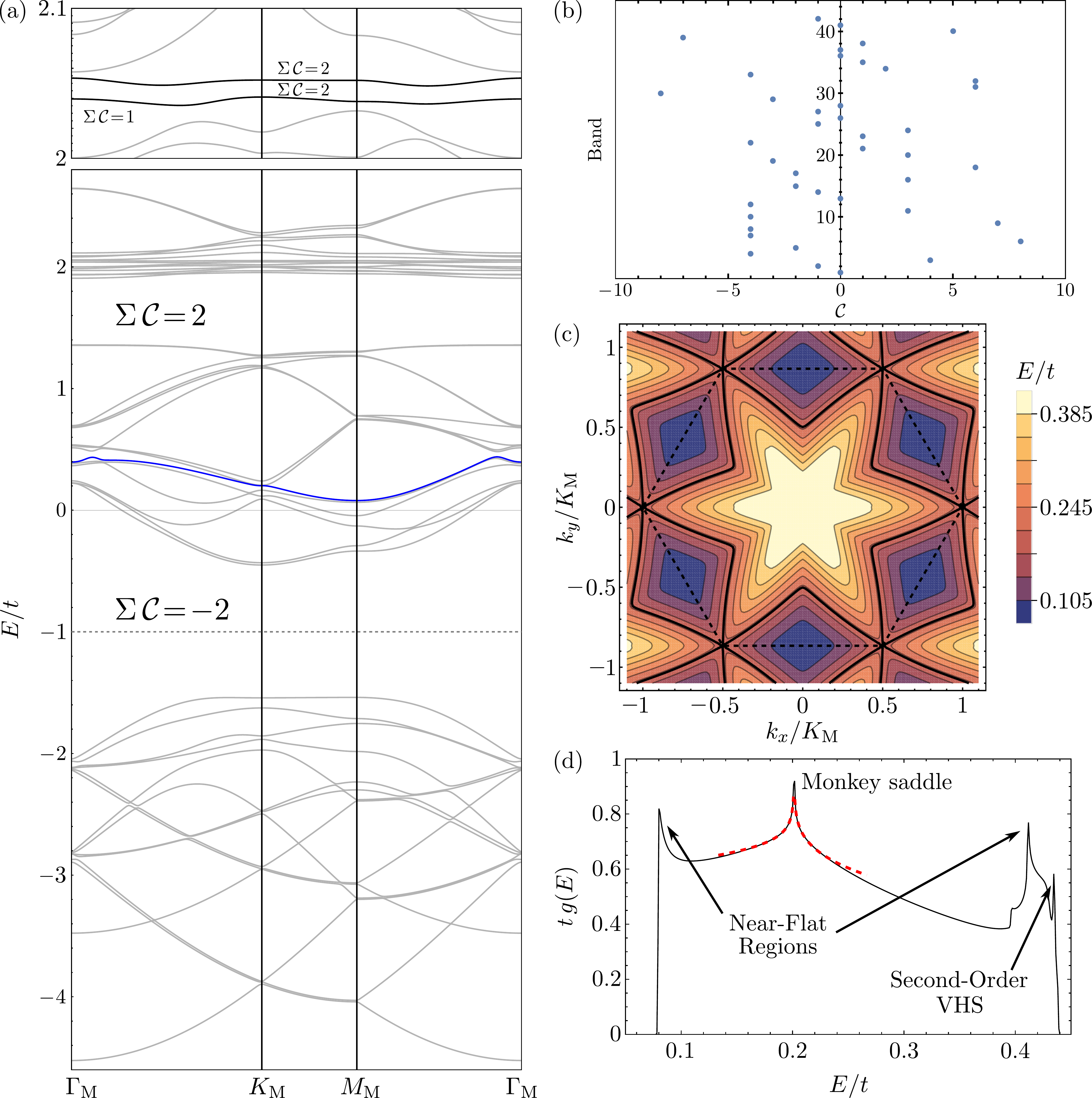}
    \caption{Band structure and Chern numbers for AA TBK tuned to a topological monkey saddle singularity, alongstide the DOS fo topological band hosting the singularity. Parameters: $\theta_{c} = 38.2^{\circ}$, $a = 0.5338$ nm, $d_{\perp} = 0.6596$ nm, $t_{\text{H}} = 0.18147 t$, $\phi = \pi/2$, $t_{\perp} = 0.3 t$, and $\gamma = 20$ \cite{Ye2018,Lima2019}. (a) Band structure with the monkey saddle host band highlighted (band 20). The sums of the Chern numbers in the insulating gaps are indicated. The top panel focuses on bands 35 and 36 (highlighted in black) (b) The set of Chern numbers for each band. We note that all degeneracies are lifted. Ref. \cite{SMref} provides details on checking the numerical convergence of the Chern numbers. (c) Energy contours of band 20 across the entire MBZ (boundary marked by the dashed line) with the HOVHS energy contour highlighted by the bold line. (d) Single-spin DOS per unit volume for band 20. The red dashed line marks our fitting of the monkey saddle divergence whose leading order behavior is $g(\varepsilon) \propto \varepsilon^{-1/3}$.}
    \label{Topological_HOVHS}
\end{figure}

\subsection{Two-Fold Singularities}

The appearance of singularities (either localized or delocalized) with a two-fold symmetry at the $M_{\text{M}}$ points is expected in TBK given the VHSs that exist at the BZ edges of the monolayer. We immediately note that fourth-order singularities of the form $\varepsilon \sim k_{x}^{2} - k_{y}^{4}$ can be acquired without the need for sophisticated hoppings to break inversion or time-reversal symmetry as illustrated by band 14 in Fig. \figRef{HOVHS_examples}{d}. Here we set $t_{d} = t_{\text{H}} = 0$ and tune the interlayer hopping amplitude to $t_{\perp} = 0.267 t$ to produce a DOS scaling of $g(\varepsilon) \sim D_{\pm} |\varepsilon|^{-1/4}$ at the MBZ edge. Moreover, the prefactors $D_{+}$ and $D_{-}$ for $\varepsilon > 0$ and $\varepsilon < 0$, respectively, yield a ratio of $D_{-}/D_{+} \simeq \sqrt{2}$, matching the expected value for this type of HOVHS \cite{Chandrasekaran2020}. This demonstrates that HOVHSs can be achieved in layered systems purely through twisting. We note that even with dimerization included, two-fold singularities can persist as in band 26 which hosts extremely flat singularities at its MBZ edges, yielding the second peak in the DOS plot of Fig. \figRef{HOVHS_examples}{a}, and in band 5 also yielding the second peak in the DOS in Fig. \figRef{HOVHS_examples}{c}. Two-fold singularities may also form away from the MBZ edges, as seen in Fig. \figRef{HOVHS_examples}{b} for band 18 (first DOS peak) hosting $C_{2z}$ singularities close to the $\Gamma_{\text{M}}$-$K_{\text{M}}$ high-symmetry line. Finally, we note that it may also be possible to engineer fourth-order singularities around the MBZ edges through dimerization and additional parameters beyond those considered here, which we discuss in Ref. \cite{SMref}.

\subsection{Topological Monkey Saddles}

Complex NNN hopping is an alternative way to gap out the Dirac cones that appear at MBZ corners, creating monkey saddle singularities through the breaking of $\mathcal{T}$. We show in Fig. \ref{Topological_HOVHS} the electronic structure for $AA$ TBK with a purely imaginary NNN hopping tuned to yield a delocalized monkey saddle in band 20. We find that all degeneracies are lifted to result in the majority of bands becoming topologically non-trivial, with Chern numbers as large as $|\mathcal{C}| = 8$ being observed in bands 6 and 30 (Fig. \figRef{Topological_HOVHS}{b}). Two major gaps ($\Delta E \sim 0.5 t$ to $t$) open in the band structure and split the bands into three equal sets of 14 bands (Fig. \figRef{Topological_HOVHS}{a}), with the lowest energy set giving a total Chern number of $-2$ while the middle energy set possesses a total Chern number of 4. Therefore, the $AA$ TBK system will also become a Chern insulator exhibiting topologically protected edge states yielding an intrinsic anomalous Hall conductivity of $\sigma_{yx}^{\text{int}} = \pm 2 e^{2}/h$ in these gaps. Closer inspection of the near-flat band region for this particular case reveals three more, albeit smaller, insulating gaps ($\Delta E \sim 0.01 t$) around bands 35 and 36, see Fig. \figRef{Topological_HOVHS}{a}. However, band 36 is topologically trivial and so we only observe $\sigma_{yx}^{\text{int}} = e^{2}/h$ below band 35 and $\sigma_{yx}^{\text{int}} = 2 e^{2}/h$ above bands 35 and 36. Regarding band 20, it acts as a host to a delocalized topological HOVHS with a Chern number of $\mathcal{C} = 3$. The DOS for this band (Fig. \figRef{Topological_HOVHS}{d}) reveals multiple singularities: a set of six closely situated second-order VHSs near the origin approaching a sixth-order singularity near the top of the band, and two regions where the band becomes extremely flat. The monkey saddle peak is well fitted by a DOS with a leading order correction that scales as $g(\varepsilon) \propto \varepsilon^{-1/3}$.

\begin{figure}[t]
    \centering
    \includegraphics[width=\linewidth]{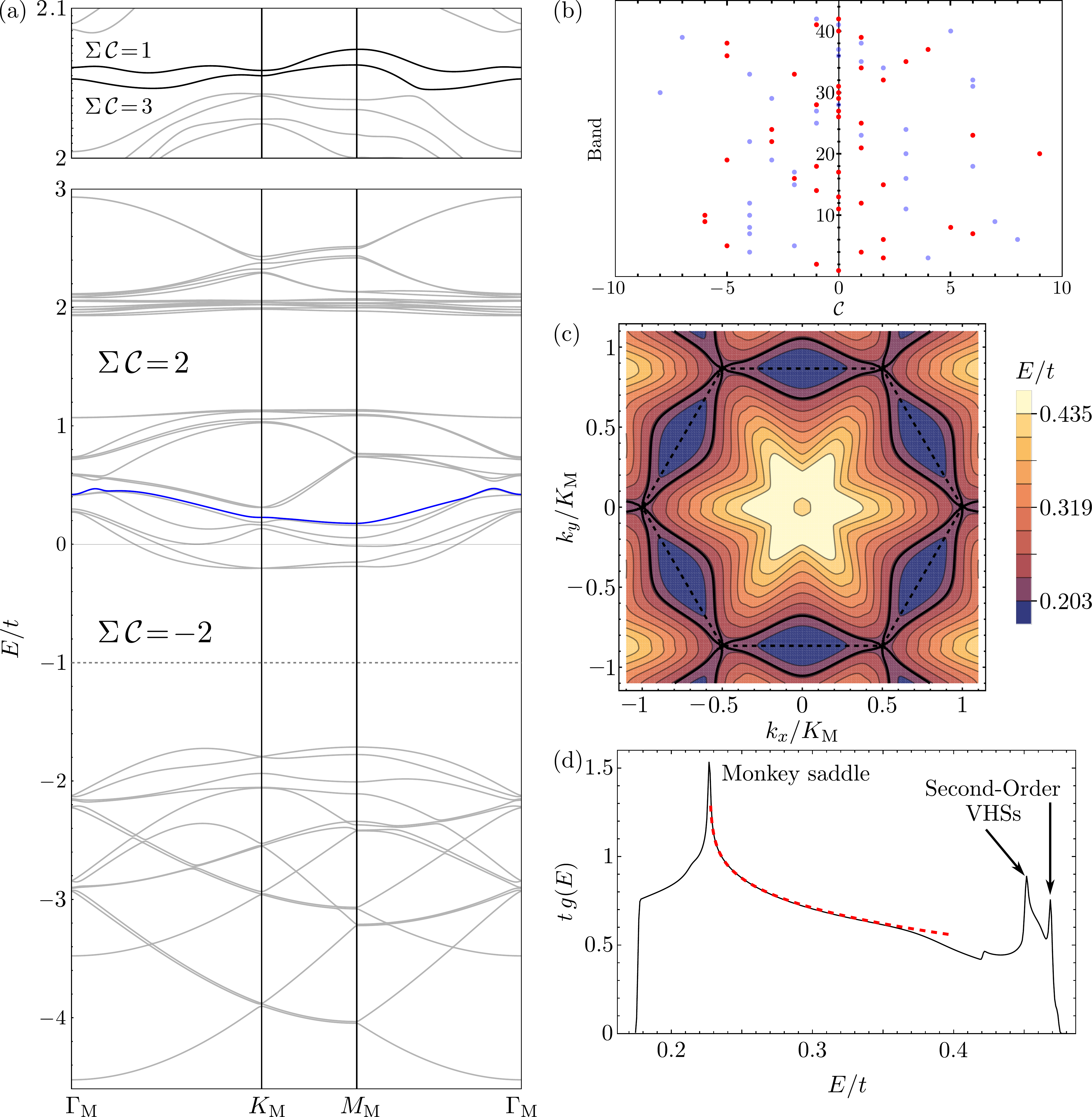}
    \caption{Band structure and Chern numbers for interlocked TBK tuned to a topological monkey saddle singularity, alongstide the DOS fo topological band hosting the singularity. Parameters: $\theta_{c} = 38.2^{\circ}$, $a = 0.5338$ nm, $d_{\perp} = 0.6596$ nm, $t_{\text{H}} = 0.26589 t$, $\phi = \pi/2$, $t_{\perp} = 0.3 t$, and $\gamma = 20$ \cite{Ye2018,Lima2019}. (a) Band structure with the monkey saddle host band highlighted (band 20). The sum of the Chern numbers for the first and second sets of 14 bands are indicated. The top panel focuses on bands 35 and 36 (highlighted in black). (b) The set of Chern numbers for each band for interlocked (red) and $AA$ TBK (pale blue) when tuned to host a monkey saddle in band 20, respectively. Again, all degeneracies are lifted. (c) Energy contours of band 20 across the entire MBZ (boundary marked by the dashed line) with the HOVHS energy contour is highlighted by the bold line. (d) Single-spin DOS per unit volume for band 20. The red dashed line marks our fitting of the monkey saddle divergence whose leading order behavior is $g(\varepsilon) \propto \varepsilon^{-1/3}$.}
    \label{Topological_HOVHS_interlocked}
\end{figure}

Changing the stacking from $AA$ to interlocked, another delocalized topological monkey saddle in band 20 is revealed with a larger choice of $t_{\text{H}}$, see Fig. \figRef{Topological_HOVHS_interlocked}. However, the topology of all bands is altered drastically in comparison to the $AA$ system with a Chern number of $\mathcal{C} = 9$ now being associated to the monkey saddle band (Fig. \figRef{Topological_HOVHS_interlocked}{b}). Moreover, the minor insulating gaps above band 36 and below band 35 remain and appear larger than in the $AA$ system. Likewise, the two major insulating gaps also persist, again splitting the bands into three groups of 14 bands and allowing for a quantum anomalous Hall response of $\sigma_{yx}^{\text{int}} = \pm 2 e^{2}/h$ equivalent to the $AA$ stacked system. In contrast, the insulating gap between bands 35 and 36 has now become an indirect gap preventing a fifth region of Chern insulator behaviour for this tuning of the electronic structure. Nonetheless, we note that it may be possible to tune the system to open this gap and encourage a giant anomalous Hall conductivity with $\sigma_{yx}^{\text{int}} = \pm 6 e^{2}/h$ due to the sum of Chern numbers from bands 35 and below giving $\Sigma \mathcal{C} = 6$. Further examples of topological monkey saddles are provided in Ref. \cite{SMref}.

\subsection{The Role of Sublattice Interference}

\begin{figure}[t]
    \centering
    \includegraphics[width=\linewidth]{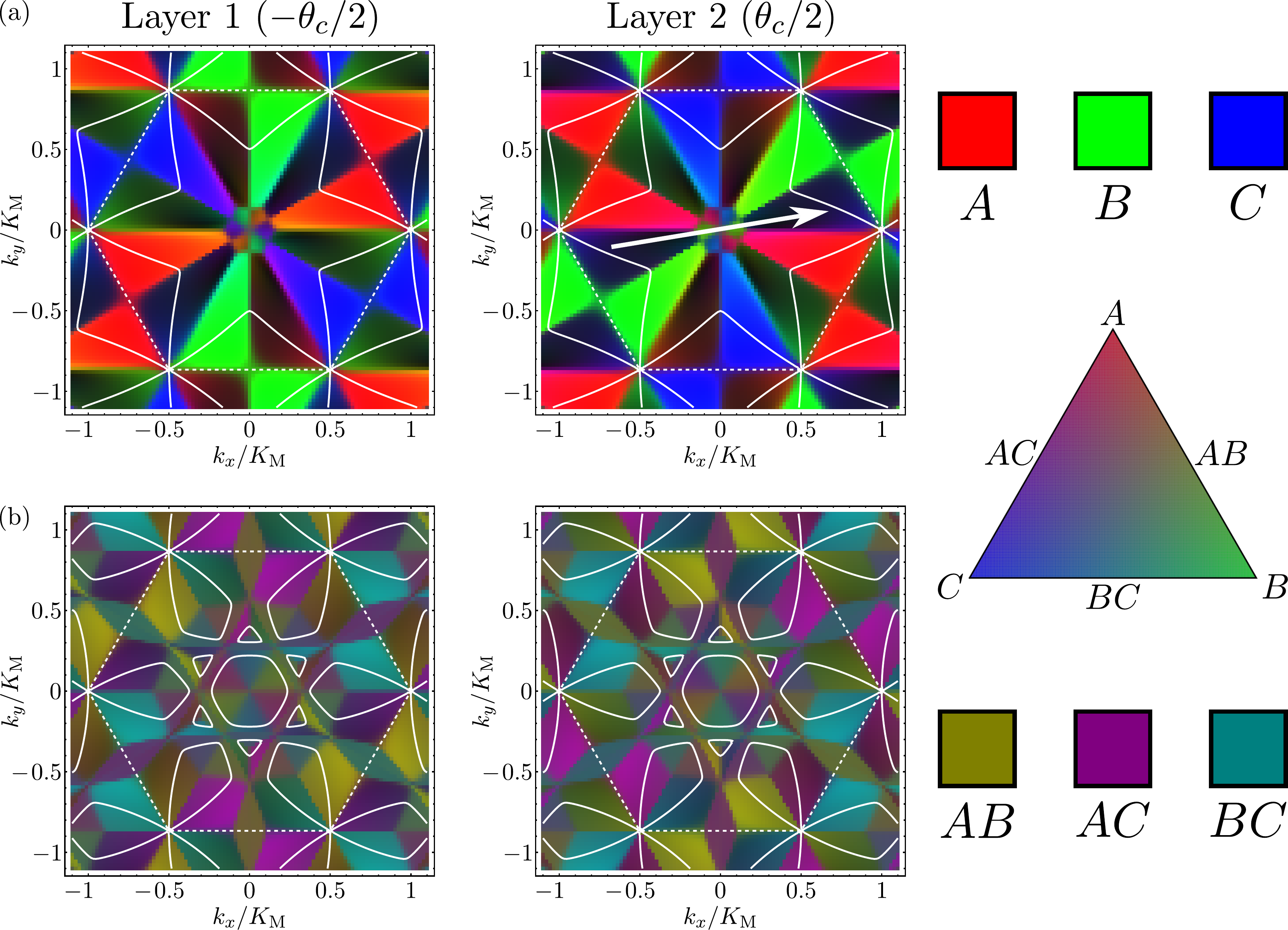}
    \caption{Sublattice polarisation for different bands tuned to host a monkey saddle singularity. Colour maps for two bands in $AA$ TBK tuned to host a monkey saddle singularities at the MBZ corners indicating their sublattice projection throughout the MBZ within the bottom (left) and top (right) layers. If the eigenstate is completely localized to a sinlge sublattice it is given a colour of red ($A$), green ($B$), or blue ($C$). However, if the state is equally localized between two sublattices, it is given by equal mixing of their respective colours. The white contours highlight the Fermi surface at the critical energy of the HOVHS and the white dashed hexagon indicates the MBZ boundary. The white arrow is an example of a possible nesting vector between Fermi surface regions related by a $\pi$ rotation. In both cases we use $\theta_{c} = 38.2^{\circ}$, $a = 0.5338$ nm, $d_{\perp} = 0.6596$ nm, $t_{\perp} = 0.3 t$, and $\gamma = 20$ \cite{Ye2018,Lima2019}. (a): Band 20 with $t_{\text{H}} = 0.18147 t$ and $\phi = \pi/2$. (b): Band 8 with $t_{d} = 0.0862 t$.}
    \label{Sublattice_interference_plots}
\end{figure}

Nesting vectors act to enhance particle-hole fluctuations by connecting large parallel sections of the Fermi surface and can play a crucial role in the formation of different phases \cite{Zervou2023}. However, sublattice interference in a Kagom\'{e} monolayer can greatly affect these vectors by requiring that they further connect sections of the Fermi surface with the same sublattice projection \cite{Kiesel2012}. Specifically, it has been shown that the critical contour associated to the VHS at 5/12 filling has six nesting vectors, as opposed to the three we might expect akin to graphene tuned to its VHS, which connect the BZ corners to the opposing edge $M$ points \cite{Kiesel2012}. This results in a preference for $f$-wave superconductivity in a Kagom\'{e} Hubbard system. However, the energy contours associated to the various HOVHSs are not simple hexagons due to their nature of curving the Fermi surface and thus naturally suppress nesting effects \cite{Nag2024arxiv}. Nonetheless, due to HOVHSs tending to enhance the transition temperatures of ordered phases, thermal broadening of the Fermi surface should be accounted for to yield \textit{diffusive nesting} vectors that may connect regions of the thermally broadened Fermi surface \cite{Beck2024arxiv}. Specifically, for $AA$ TBK, we may identify diffusive nesting vectors that connect sections of the Fermi surface that are related to each another by a $\pi$ rotation about the $z$-axis, see Fig. \figRef{Sublattice_interference_plots}{a}. The precise diffusive nesting vectors depend on the filling and the associated shape of the Fermi surface, but from the symmetry point of view, it is possible for them to survive. The symmetries of the Fermi surface respect at least those of the BZ as defined here.

By projecting the eigenstates for TBK onto the $A$, $B$, and $C$ sublattices of the first and second layers, we observe a similar sublattice polarisation pattern across the entire MBZ that matches with that seen for monolayer Kagom\'{e} in Ref. \cite{Kiesel2012}. Specifically, the Fermi surface sections that possess the same sublattice polarisation form a $C_{2z}$ symmetry, see Fig. \ref{Sublattice_interference_plots}, thus matching the natural nesting vectors that can be inferred directly from the Fermi surface contour. This would suggest that sublattice interference around the full Fermi surface is unlikely to have as large an effect as in the monolayer for these cases. However, other singularities exist within the TBK band structure with vastly different Fermi surface shapes, meaning we cannot rule out sublattice interference as a means to restrict the nesting vectors beyond the requirements of the Fermi surface geometry. Moreover, sublattice interference may play a significant role in constructing the effective interaction channels through connecting pairs of electrons situated close to a HOVHSs \cite{Classen2020,Lee2024}. The sublattice projections of the bands reveal that the eigenstates can either be localized primarily on a single sublattice or between two sublattices. Band 20 in Fig. \figRef{Sublattice_interference_plots}{a} shows the former case while band 8 in Fig. \figRef{Sublattice_interference_plots}{b} exhibit the latter with admixing between sublattices. Finally, we note that the dihedral symmetry along the $x$-axis is clearly illustrated by the sublattice projections of the two layers in Fig. \ref{Sublattice_interference_plots}, with $A$ and $A'$ mapping onto one another, while $B$ ($C$) maps onto $C'$ ($B'$) for the $\theta_{c} = 38.2^{\circ}$ $AA$ TBK system, see Fig. \ref{MUC_examples} in Appendix \ref{Monkey_saddle_details_sec} for the MUCs. In both cases, the sublattice polarisation around the HOVHSs is highly non-trivial, so we leave interaction considerations and the interplay of HOVHSs and sublattice intereference to future works.

\section{Discussion}

In this article, we have presented a detailed analysis of the symmetries and band structure for twisted bilayer Kagom\'{e} vdW heterostuctures for a large commensurate twist angle and identified its potential to host exotic topological HOVHS. Specifically, we encountered three types of Van Hove singularity in the band structure: monkey saddles around the $K_{\text{M}}$ point, six-fold singularities around $\Gamma_{\text{M}}$, and two-fold singularities at the $M_{\text{M}}$ point. These singularities could be achieved through the use of dimerization, an applied out-of-plane electric field, and complex NNN hopping. The monkey saddle singularities were found to occur in several bands and for a large range of tuning parameters, making them relatively accessible as they can be expected to occur in a plethora of TBK systems. In studying this type of singularity, we classified them into two categories: delocalized and localized. Both types of singularity can be expected to have significant effects on the electric and thermal transport coefficients, shifting them from the typical Fermi liquid behavior. Through complex NNN hopping, we also found them to emerge in topological bands, creating topological HOVHSs. Moreover, we observed the emergence of five insulating gaps enabling quantum anomalous Hall behaviour. This suggests that TBK hosts a rich landscape of transport phenomena ranging from non-Fermi liquid behavior to topologically protected edge states. Furthermore, we found that the TBK system hosted bands with large Chern numbers with a magnitude of order 10 when tuned to host a HOVHS. This is of particular interest in searching for systems with a giant quantum anomalous Hall response and systems hosting a large number of active edge states. Our work therefore serves as a demonstration that such strong topological signatures may be achievable in twisted heterostructures.

Symmetry analysis identified the combination of dihedral rotation and time reversal, $\mathcal{T} C_{2}'$, to play a central role in determining the number of possible HOVHS that may occur simultaneously due to the breaking of time-reversal symmetry changing the symmetry from a trivial lattice point group to a non-trivial magnetic point group. The application of an out-of-plane electric field acts to break this symmetry and reduce the number of HOVHSs. We expect that the $AA$ and interlocked stackings will have similar properties and phase diagrams when they possess the same symmetries. This is not the case when Haldane tunneling is included while dimerization is absent. In contrast, the $AB$ system is likely to exhibit the most unique phase diagram because it only hosts singularities with $C_{2}$ symmetry. If we consider a scenario where $AB$ TBK is in the same phase as the $AA$ and interlocked systems, then we would still expect unique signatures in the transport coefficients and electron self-energy courtesy of the different types of HOVHS \cite{Mousatov2020,Stangier2022}.

The results presented here act as a guide to how topological HOVHSs can be achieved when constructing vdW heterostructures with Kagom\'{e} patterns. Currently, there are several candidate systems in which to probe these ideas: the largest class of candidate materials are MOFs, but recently N-doped Kagom\'{e} graphene and 2D $\pi$-conjugated polymers have been synthesised, realising Kagom\'{e} patterns formed predominantly of triangular carbon atom islands \cite{Galeotti2020,Palwak2021}. There is intensive research for more materials of Kagom\'{e} structure where the results of this work are relevant. For the case of small twist angles, whilst the treatment of a breathing Kagom\'{e} lattice as a dimerized lattice is no longer applicable, the symmetry analysis of the HOVHSs presented here remains true in general for all commensurate angles. Additionally, the two types of HOVHSs identified here, localised and delocalised, should also appear at small twist angles for both commensurate and incommensurate angles handled via Bistritzer-MacDonald type analysis. Such low-energy continuum models would introduce topological HOVHSs characterised by valley Chern numbers instead, which may lead to an interesting interplay between valley-polarized phases and topology. Regarding other lattices, the same symmetries occur in hexagonal and triangular lattices. The dense flat band region of the Kagom\'{e} bilayer though, which led to the rich structure of HOVHSs is the reason that we took this lattice as an example. Future work will focus on the role of interactions in the TBK system to discover how the flat bands and the plethora of topological HOVHSs influence phase formation.

\acknowledgements{}
We thank Claudio Chamon, Andrey Chubukov, Na\"{e}mi Leo, Natalia Perkins and Ioannis Rousochatzakis for useful discussions, and especially Titus Neupert and Luiz Santos for reading the manuscript and providing insightful comments and questions. The work has been supported by the EPSRC grants EP/T034351/1 and EP/X012557/1.

\appendix

\section{Two Classes of Moir\'{e} Superlattices} \label{Moire_class_sec}

\begin{figure*}[t]
    \centering
    \includegraphics[width=0.85\textwidth]{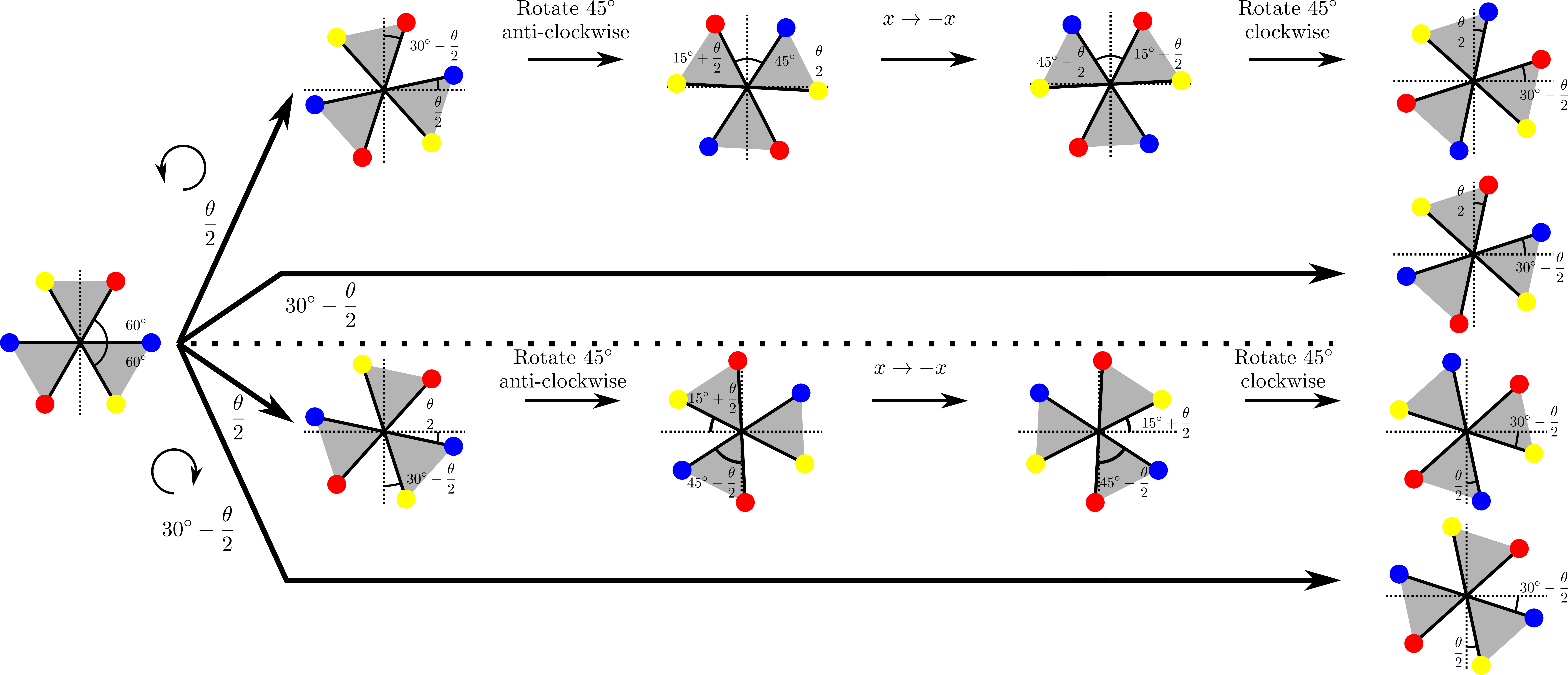}
    \caption{Schematic representation for the relation of layers rotated by an arbitrary angle of $\theta/2$ and $\theta'/2=\pi/6 - \theta/2$ by reflection in the line of $y = x$. We start with the hexagon centre of a Kagom\'{e} lattice at the origin and label the three sublattices using red, blue, and yellow coloured dots. We connect these to the origin and shade down triangles while leaving up triangles blank. The first line shows the hexagon rotated anti-clockwise by $\theta/2$ and then transformed by reflection in $y =x$. The second line shows the hexagon rotated anti-clockwise by $\theta'/2$. Similarly, the third line shows the hexagon rotated clockwise by $\theta/2$ and then transformed by reflection in $y =x$, while the fourth line shows the hexagon rotated clockwise by $\theta'/2$.}
    \label{Moire_classes_relation_schematic}
\end{figure*}

The set of commensurate angles are defined by a pair of unique coprime integers $(m,n)$, such that $m > n > 0$. The commensurate angle $\theta_{c}$ associated to $(m,n)$ is given by eq. \ref{Commensurate_angle_definition}. These commensurate angles can be split into two classes: those with $n$ divisible by 3 and those with $n$ not divisible by 3. If $\theta_{c}$ is defined by a choice of $n$ that is divisible by 3, then it has a partner commensurate angle that is associated to a choice of $n$ that is not divisible by 3, $\theta_{c}' = \pi/3 - \theta_{c}$ \cite{Scheer2022}. When constructing a commensurate twisted bilayer we may choose to work in the twist-symmetric frame where layer 1 is rotated clockwise by $\theta_{c}/2$ and layer 2 is rotated anti-clockwise by $\theta_{c}/2$. We can relate the two classes of superlattices in the yielded by $\theta_{c}$ and $\theta_{c}'$ in the twist-symmetric frame by reflection in the line of $y = x$, as we prove below.

Let us start by considering a Kagom\'{e} hexagon centred on the origin and label each of the sublattices around the corners, see the first line of Fig. \ref{Moire_classes_relation_schematic}. We can choose to rotate this layer anti-clockwise by $\theta_{c}/2$ to acquire one of its orientations in the twist-symmetric frame for one of the superlattice classes. To reflect this in the line of $y = x$, we rotate the who system anti-clockwise by $45^{\circ}$, reflect it in the $y$-axis ($x \rightarrow -x$), and then rotate the system clockwise by $45^{\circ}$, as illustrated in Fig. \ref{Moire_classes_relation_schematic}. The other superlattice class would be given by choosing to rotate this layer by $\theta_{c}'/2 = (\pi/3 - \theta_{c})/2$ anti-clockwise, see the second line of Fig. \ref{Moire_classes_relation_schematic}. By comparing the first and second lines of Fig. \ref{Moire_classes_relation_schematic}, we immediately see that the $\theta_{c}$ moir\'{e} pattern is mapped onto the $\theta_{c}'$ moir\'{e} pattern with a swapping of sublattices. From the perspective of geometry there is no difference, but for the Hamiltonian this can have consequences due to the relative orientation of the sublattices determining the topological contribution.

Alternatively, we can instead choose to rotate this layer clockwise by either $\theta_{c}/2$ or $\theta_{c}'/2$ to consider the other layer in the twisted bilayer, as shown in the third and fourth lines of Fig. \ref{Moire_classes_relation_schematic}, respectively. By transforming $\theta_{c}/2$ layer by reflection in line of $y = x$, see the third line of Fig. \ref{Moire_classes_relation_schematic}, we again acquire the same geometry as the $\theta_{c}'/2$ layer. However, unlike the anti-clockwise rotation, not only is there a swapping of sublattices, but the orientation of up and down triangles is also swapped, as shown by the switching of highlighted and blank regions in Fig. \ref{Moire_classes_relation_schematic}. Therefore, while the geometry of the two classes are connected by a simple reflection, the topological aspect of the Hamiltonian we once again by flipped and the nature of dimersation will also be reversed.

\section{Derivation of the TBK Tight-Binding Hamiltonian} \label{Hamiltonian_derivation_sec}

\begin{figure}[t]
    \centering
    \includegraphics[width=0.8\linewidth]{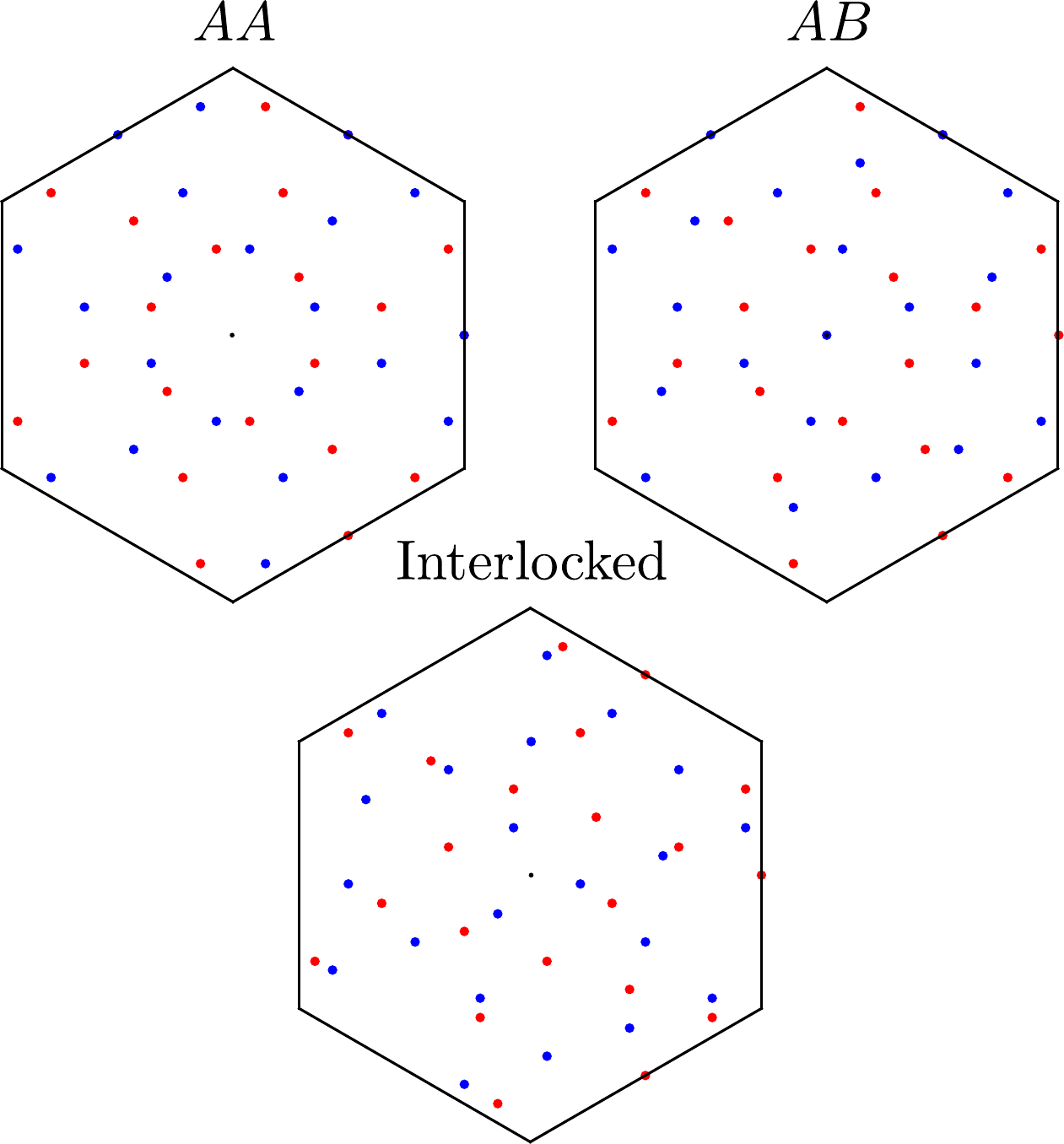}
    \caption{Wigner-Seitz constructed moir\'{e} unit cells for the three possible stacking choices with a $\theta = 38.2^{\circ}$ twist applied about the hexagon centre of layer 1. The red and blue dots belong to layers 1 and 2, respectively, while the small black dot indicates the twist origin and Wigner-Seitz lattice site. We show the MUCs in the twist symmetric frame.}
    \label{MUC_examples}
\end{figure}

Let us start by considering a moir\'{e} unit cell due to an arbitrary commensurate twist angle of $\theta_{c}$. Each monolayer contributes $N^{\text{M}}$ triangular lattice sites to the unit cell, and hence $3N^{\text{M}}$ sublattice sites, thus leaving us with $6N^{\text{M}}$ sublattice sites within the moir\'{e} unit cell. We may then introduce $c_{il\alpha}^{\dagger}$ $(c_{il\alpha}^{\null})$ creation (annihilation) operators to describe the addition (removal) of an electron from sublattice $\alpha$ of layer $l$ in the moir\'{e} unit cell centred on $\mathbf{R}_{i}$, where $\{\alpha \, | \, \alpha \in \mathbb{Z}, \, 1 \leq \alpha \leq 3N^{\text{M}} \}$. We prescribe $\boldsymbol{\delta}_{\alpha}^{(l)}$ as the position of sublattice site $\alpha$ in layer $l$ within the moir\'{e} unit cell. The general Hamiltonian for this system may then be written as
\begin{widetext}
\begin{equation}
\begin{split}
    H = - \sum_{i} \sum_{l} \sum_{\alpha \neq \beta} t_{l,ii}^{\alpha\beta} c_{il\alpha}^{\dagger} c_{il\beta}^{\null}
    &- \sum_{\substack{\langle i,j \rangle, \\ \langle\!\langle i,j \rangle\!\rangle}} \sum_{l} \sum_{\alpha,\beta} t_{l,ij}^{\alpha\beta} c_{il\alpha}^{\dagger} c_{jl\beta}^{\null}
    \\
    &- \sum_{i} \sum_{\alpha,\beta} t_{\perp,ii}^{\alpha\beta} (c_{i1\alpha}^{\dagger} c_{i2\beta}^{\null} + c_{i2\beta}^{\dagger} c_{i1\alpha}^{\null})
    - \sum_{\substack{\langle i,j \rangle, \\ \langle\!\langle i,j \rangle\!\rangle}} \sum_{\alpha,\beta} t_{\perp,ij}^{\alpha\beta} (c_{i1\alpha}^{\dagger} c_{j2\beta}^{\null} + c_{j2\beta}^{\dagger} c_{i1\alpha}^{\null} ),
\end{split}
\end{equation}
assuming that the tunneling is real and decays sufficiently quickly to only allow tunneling up to NNN moir\'{e} unit cells. The first and second lines represent intralayer tunneling while the third and fourth lines account for the interlayer tunneling. In writing this Hamiltonian, we noted that the tunneling energy should not depend on the direction of tunneling, $t_{\perp,ij}^{\alpha\beta} = t_{\perp,ji}^{\beta\alpha}$. Furthermore, we define the MUC as the Wigner-Seitz constructed unit cell centred on the twist origin, see Fig. \ref{MUC_examples}, and use $\langle i,j \rangle$ and $\langle\!\langle i,j \rangle\!\rangle$ to denote NN and NNN unit cells, respectively.

For simplicity, let us assume that the tunneling elements may be written as $t_{l,ij}^{\alpha\beta} = t^{\alpha\beta} f(|\mathbf{R}_{i}^{\null} + \boldsymbol{\delta}_{\alpha}^{(l)} - \mathbf{R}_{j}^{\null} - \boldsymbol{\delta}_{\beta}^{(l)}|)$ and $t_{\perp,ij}^{\alpha\beta} = t_{\perp0}^{\alpha\beta} f_{\perp}^{\null}(|\mathbf{R}_{i}^{\null} + \boldsymbol{\delta}_{\alpha}^{(1)} - \mathbf{R}_{j}^{\null} - \boldsymbol{\delta}_{\beta}^{(2)}|)$, with $t^{\alpha\beta} = t^{\beta\alpha}$ and $t_{\perp0}^{\alpha\beta} = t_{\perp0}^{\beta\alpha}$ (the superscripts remain to allow different orbital overlaps between sublattices), where $f(r)$ and $f_{\perp}^{\null}(r)$ are functions that decay as $r$ increases. Now we can move to momentum space,
\begin{equation}
\begin{split}
    H = &- \sum_{\mathbf{k}} \sum_{l} \sum_{\alpha \neq \beta} t^{\alpha\beta} f(|\boldsymbol{\delta}_{\beta\alpha}^{(l)}|) e^{i\mathbf{k} \cdot \boldsymbol{\delta}_{\beta\alpha}^{(l)}} c_{\mathbf{k}l\alpha}^{\dagger} c_{\mathbf{k}l\beta}^{\null}
    - \sum_{\mathbf{k}} \sum_{l} \sum_{\alpha,\beta} t^{\alpha\beta} \widetilde{\Gamma}_{l}^{\alpha\beta}(\mathbf{k}) c_{\mathbf{k}l\alpha}^{\dagger} c_{\mathbf{k}l\beta}^{\null} \\
    &- \sum_{\mathbf{k}} \sum_{\alpha,\beta} t_{\perp0}^{\alpha\beta} \left[ \left( f_{\perp}^{\null}(|\boldsymbol{\delta}_{\beta\alpha}^{(21)}|) e^{i\mathbf{k} \cdot \boldsymbol{\delta}_{\beta\alpha}^{(21)}} + \widetilde{\Gamma}_{\perp}^{\alpha\beta}(\mathbf{k}) \right) c_{\mathbf{k}1\alpha}^{\dagger} c_{\mathbf{k}2\beta}^{\null} + \left( f_{\perp}^{\null}(|\boldsymbol{\delta}_{\beta\alpha}^{(21)}|) e^{-i\mathbf{k} \cdot \boldsymbol{\delta}_{\beta\alpha}^{(21)}} + \widetilde{\Gamma}_{\perp}^{\alpha\beta}(\mathbf{k})^{*}\right) c_{\mathbf{k}2\beta}^{\dagger} c_{\mathbf{k}1\alpha}^{\null} \right],
    \label{General_k_space_NNN_Hamiltonian}
\end{split}
\end{equation}
\end{widetext}
with $\boldsymbol{\delta}_{\beta\alpha}^{(l)} = \boldsymbol{\delta}_{\beta}^{(l)} - \boldsymbol{\delta}_{\alpha}^{(l)}$, $\boldsymbol{\delta}_{\beta\alpha}^{(21)} = \boldsymbol{\delta}_{\beta}^{(2)} - \boldsymbol{\delta}_{\alpha}^{(1)}$,
\begin{equation}
\begin{split}
    \widetilde{\Gamma}_{l}^{\alpha\beta}(\mathbf{k}) &= \sum_{i=1}^{18} f(|\boldsymbol{\delta}_{\beta\alpha}^{(l)} + \mathbf{c}_{i}^{\text{M}}|) e^{i \mathbf{k} \cdot (\boldsymbol{\delta}_{\beta\alpha}^{(l)} + \mathbf{c}_{i}^{\text{M}})},
    \\
    \widetilde{\Gamma}_{\perp}^{\alpha\beta}(\mathbf{k}) &= \sum_{i=1}^{18} f_{\perp}(|\boldsymbol{\delta}_{\beta\alpha}^{(21)} + \mathbf{c}_{i}^{\text{M}}|) e^{i \mathbf{k} \cdot (\boldsymbol{\delta}_{\beta\alpha}^{(21)} + \mathbf{c}_{i}^{\text{M}})},
\end{split}
\end{equation}
and
\begin{equation}
\begin{split}
    \{\mathbf{c}_{i}^{\text{M}}\} &= \{ \pm \mathbf{a}_{1}^{\text{M}}, \pm \mathbf{a}_{2}^{\text{M}}, \pm (\mathbf{a}_{2}^{\text{M}} - \mathbf{a}_{1}^{\text{M}}),
    \\
    &\!\!\pm 2 \mathbf{a}_{1}^{\text{M}}, \pm 2 \mathbf{a}_{2}^{\text{M}}, \pm 2 (\mathbf{a}_{2}^{\text{M}} - \mathbf{a}_{1}^{\text{M}}),
    \\
    &\!\!\pm (\mathbf{a}_{2}^{\text{M}} + \mathbf{a}_{1}^{\text{M}}), \pm (2 \mathbf{a}_{2}^{\text{M}} - \mathbf{a}_{1}^{\text{M}}), \pm (2 \mathbf{a}_{1}^{\text{M}} - \mathbf{a}_{2}^{\text{M}}) \},
\end{split}
\end{equation}
being the set of moir\'{e} lattice vectors connecting NN and NNN moir\'{e} unit cells. We can further simplify the Hamiltonian in eq. \ref{General_k_space_NNN_Hamiltonian} by introducing
\begin{equation}
\begin{split}
    \Gamma_{l}^{\alpha\beta}(\mathbf{k}) &= f(|\boldsymbol{\delta}_{\beta\alpha}^{(l)}|) e^{i\mathbf{k} \cdot \boldsymbol{\delta}_{\beta\alpha}^{(l)}} (1 - \delta_{\alpha\beta}) + \widetilde{\Gamma}_{l}^{\alpha\beta}(\mathbf{k}),
    \\
    \Gamma_{\perp}^{\alpha\beta}(\mathbf{k}) &= f_{\perp}(|\boldsymbol{\delta}_{\beta\alpha}^{(21)}|) e^{i\mathbf{k} \cdot \boldsymbol{\delta}_{\beta\alpha}^{(21)}} + \widetilde{\Gamma}_{\perp}^{\alpha\beta}(\mathbf{k}),
\end{split}
\end{equation}
which yields
\begin{equation}
\begin{split}
    H &= - \sum_{\mathbf{k}} \sum_{\alpha,\beta} \Big( \sum_{l} t^{\alpha\beta} \Gamma_{l}^{\alpha\beta}(\mathbf{k}) c_{\mathbf{k}l\alpha}^{\dagger} c_{\mathbf{k}l\beta}^{\null}
    \\
    &+ t_{\perp0}^{\alpha\beta} \left[ \Gamma_{\perp}^{\alpha\beta}(\mathbf{k}) c_{\mathbf{k}1\alpha}^{\dagger} c_{\mathbf{k}2\beta}^{\null} + \Gamma_{\perp}^{\alpha\beta}(\mathbf{k})^{*} c_{\mathbf{k}2\beta}^{\dagger} c_{\mathbf{k}1\alpha}^{\null} \right] \Big).
\end{split}
\end{equation}

We include complex NNN tunneling as a purely intralayer process. Written in real space, this mechanism contributes
\begin{equation}
\begin{split}
    H_{\text{H}} = &- t_{\text{H}}^{\null} \sum_{l} \sum_{i} \sideset{}{''}\sum_{\alpha\neq\beta} e^{i\mathcal{S}_{\alpha\beta} \phi} c_{il\alpha}^{\dagger} c_{il\beta}^{\null}
    \\
    &- t_{\text{H}}^{\null} \sum_{l} \sum_{\langle i,j \rangle} \sideset{}{''}\sum_{\alpha,\beta} e^{i\mathcal{S}_{\alpha\beta} \phi} c_{il\alpha}^{\dagger} c_{jl\beta}^{\null}
\end{split}
\end{equation}
to the Hamiltonian, where the double primes indicate the restriction of the sum to NNN sites. Moving to momentum space yields
\begin{equation}
    H_{\text{H}}^{\null} = - t_{\text{H}}^{\null} \sum_{\mathbf{k}} \sum_{l} \sum_{\alpha,\beta} \Gamma_{\text{H}}^{\alpha\beta}(\mathbf{k}) e^{i \mathcal{S}_{\alpha\beta}\phi} c_{\mathbf{k}l\alpha}^{\dagger}c_{\mathbf{k}l\beta}^{\null},
    \label{Compact_TBK_Haldane_term}
\end{equation}
with
\begin{equation}
    \Gamma_{\text{H}}^{\alpha\beta}(\mathbf{k}) = \left[ \delta_{\langle\!\langle \alpha,\beta \rangle\!\rangle}^{\null} + \sum_{i=1}^{6} \delta_{\langle\!\langle \alpha,\beta \rangle\!\rangle}^{i} e^{i\mathbf{k} \cdot \mathbf{c}_{i}^{\text{M}}}\right],
\end{equation}
where $\delta_{\langle\!\langle \alpha,\beta \rangle\!\rangle}^{\null}$ is unity when $\alpha$ and $\beta$ are NNN within the same unit cell and zero otherwise. Similarly, $\delta_{\langle\!\langle \alpha,\beta \rangle\!\rangle}^{i}$ is unity when $\alpha$ and $\beta$ are NNN while located in NN unit cells, such that their separation is $\boldsymbol{\delta}_{\alpha}^{\null} - \boldsymbol{\delta}_{\beta}^{\null} - \mathbf{c}_{i}^{\text{M}}$, and zero otherwise.

\section{Obtaining Monkey Saddle Singularities} \label{Monkey_saddle_details_sec}

The presence of a monkey saddle singularity can be determined by checking the vanishing of all second-order terms in the expansion of the band energy about the $K_{\text{M}}$ point. Expanding a given band around the $K_{\text{M}}$ point that hosts a critical point yields
\begin{equation}
\begin{split}
    \varepsilon &= \varepsilon_{0}^{\null} + c_{0}^{(2)} p_{x}^{2} + c_{1}^{(2)} p_{x} p_{y} + c_{2}^{(2)} p_{y}^{2}
    \\
    &+ c_{0}^{(3)} p_{x}^{3} + c_{1}^{(3)} p_{x}^{2}p_{y}^{\null} + c_{2}^{(3)} p_{x}^{\null}p_{y}^{2} + c_{3}^{(3)} p_{y}^{3} + \mathcal{O}(p^{4})
\end{split}
\end{equation}
for the local dispersion where $\mathbf{p} = \mathbf{k} - \mathbf{K}_{\text{M}}$, $c_{i}^{(j)}$ are generic expansion constants, and $\varepsilon_{0}$ is the band's energy at the $K_{\text{M}}$ point. A regular VHS will manifest when the second-order coefficients are non-zero. The point group symmetry of $AA$ TBK is $D_{6}$ while for interlocked TBK it is $D_{3}$ \cite{Perkins2025}. When dimerization is added they are both $D_{3}$. If an out-of-plane electric field is applied to induce an interlayer potential, then the dihedral symmetry is broken due to one layer acquiring an on-site energy of $\Delta/2$ and the other $-\Delta/2$, reducing the point group symmetry to $C_{3z}$. The three-fold rotational symmetry of this system ensures that the MBZ corners must also exhibit $C_{3z}$ symmetry, requiring $c_{0}^{(2)} = c_{2}^{(2)}$, $c_{2}^{(3)} = -3 c_{0}^{(3)}$, $c_{1}^{(3)} = -3 c_{3}^{(3)}$, and $c_{1}^{(2)} = 0$, to yield
\begin{equation}
\begin{split}
    \varepsilon = \varepsilon_{0}^{\null} &+ c_{0}^{(2)} p^{2}
    \\
    &+ p^{3} [ c_{1}^{(3)} \cos(3\varphi) - c_{4}^{(3)} \sin(3\varphi)] + \mathcal{O}(p^{4}),
\end{split}
\end{equation}
where $\varphi$ is the local azimuthal angle.

\begin{figure}[t]
    \centering
    \includegraphics[width=\linewidth]{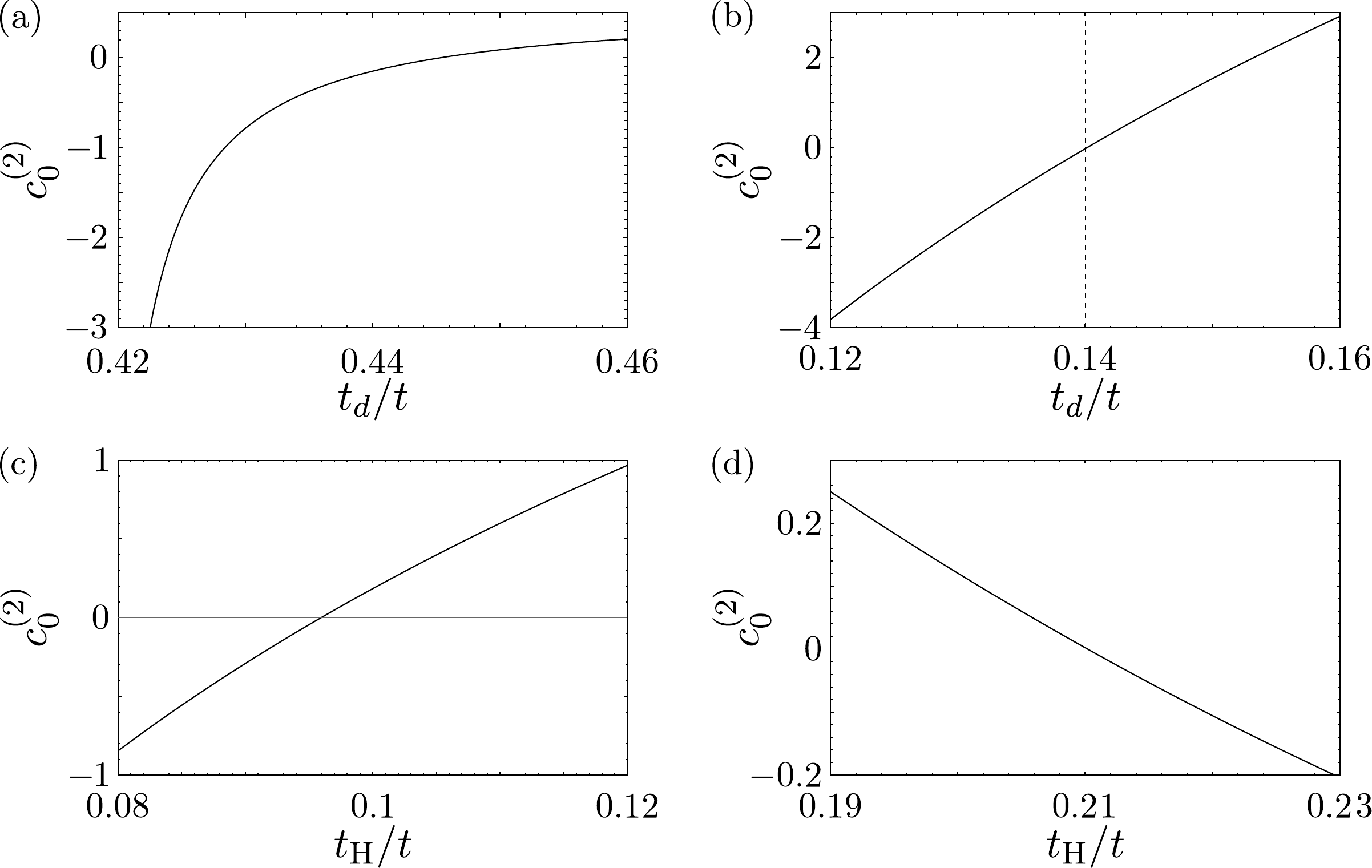}
    \caption{Variation of the second-order coefficient, $c_{0}^{(2)}$, with dimerization and Haldane hopping. (a): band 26 with $AA$ stacking. (b): band 8 with interlocked stacking. (c): band 18 with $AA$ stacking. (d): band 12 with interlocked stacking.}
    \label{Coeff_variation}
\end{figure}

Let us now restrict the above scenario to one where dihedral symmetry is preserved along the $x$-axis (i.e. no interlayer potential) with a choice of $\theta_{c}$ such that $n/3 \in \mathbb{Z}$. In this case, the dihedral symmetry manifests as a reflection $k_{y} \rightarrow -k_{y}$ symmetry in momentum space within the twist symmetric frame courtesy of the system's effective 2D nature: the dihedral symmetry can be thought of as a reflection in the $x$-axis ($y \rightarrow -y$) followed by a reflection in the $xy$ plane ($z \rightarrow -z$ i.e. swapping the layers). The out-of-plane reflection has no effect on the dispersion since the layers are identical. Hence, $C_{2x}$ dihedral symmetry sets $c_{4}^{(3)} = 0$, resulting in
\begin{equation}
    \varepsilon = \varepsilon_{0}^{\null} + c_{0}^{(2)} p^{2} + c_{1}^{(3)} p^{3} \cos(3\varphi) + \mathcal{O}(p^{4}).
\end{equation}
This effective $k_{y} \rightarrow -k_{y}$ symmetry is seen in all cases of Fig. \ref{HOVHS_examples}, Fig. S3 of Ref. \cite{SMref}, aside from Fig. S3c, where an interlayer potential is present. Fig. S3c demonstrates how this effective reflection symmetry is broken through the application of an out-of-plane electric field. If there were instead a dihedral symmetry along the $y$-axis through a choice of $n$ such that $n/3 \notin \mathbb{Z}$, then symmetry would instead require $c_{1}^{(3)} = 0$ and allow for $c_{4}^{(3)} \neq 0$.

Nonetheless, whether dihedral symmetry is preserved or not, we can demonstrate the existence of an exact monkey saddle simply by studying how $c_{0}^{(2)}$ changes as the tuning parameters are varied. If $c_{0}^{(2)} = 0$ at some point while some third-order coefficients are non-zero, then that choice of parameters will yield a monkey saddle singularity. We present the dependence of $c_{0}^{(2)}$ on the dimerization and Haldane hopping in Fig. \ref{Coeff_variation} to illustrate this in various bands of $AA$ and interlocked TBK.

\subsection{Higher-Order Corrections to the DOS} \label{Higher_order_corrections_sec}

The DOS for a band with dispersion $E(\mathbf{k})$ is given by
\begin{equation}
    g(\varepsilon) = \int_{\text{BZ}} \frac{d^2k}{(2 \pi)^2} \, \delta (E(\mathbf{k}) - \varepsilon),
\end{equation}
where the integral is over the BZ. If $E(\mathbf{k})$ has the form of a pure HOVHS, we can obtain the power-law behavior of the DOS by scaling $k_x \rightarrow |\varepsilon|^{\alpha} \, p_x$ and $k_y \rightarrow |\varepsilon |^{\beta} \, p_y$ for some appropriate $\alpha$ and $\beta$ in this integral. For example, for the HOVHS $k_x^4 - k_y^2$, we need to set $\alpha = 1/4, \beta = 1/2$ so that
\begin{align}
    g(\varepsilon) & \approx \int_{\mathbb{R}^{2}} \frac{d^2k}{(2 \pi)^2} \, \delta ( k_x^4 - k_y^2 - \varepsilon ) \nonumber \\
    & = \int_{\mathbb{R}^{2}} \frac{|\varepsilon|^{\frac{1}{4}+\frac{1}{2}} \,d^2 p}{(2 \pi)^2} \, \delta ( |\varepsilon| p_x^4 - |\varepsilon| p_y^2 - \varepsilon ) \nonumber \\
    & = |\varepsilon|^{-\frac{1}{4}} \int_{\mathbb{R}^{2}} \frac{d^2 p}{(2 \pi)^2} \, \delta ( p_x^4 - p_y^2 - \text{sign} (\varepsilon) ) \nonumber \\
    & = D_{\pm} |\varepsilon|^{-\frac{1}{4}},
\end{align}
where we have extended the integral over the entire real plane (introducing a finite error in comparison to the divergence of the DOS) and used the scaling property of the delta function: $\delta(a x) = \delta(x) / |a|$. The final integral is independent of the magnitude of $\varepsilon$ and depends only on the sign, evaluating respectively to $D_+$ for $\varepsilon > 0$ and $D_-$ for $\varepsilon < 0$. The ratio $D_{+} / D_{-}$ is a characteristic feature of each HOVHS, alongside the exponent of the power-law DOS. In particular, the power-law DOS for the $k_x^4 - k_y^2$ singularity is $|\varepsilon|^{-1/4}$. When there are higher-order terms that correct the HOVHS (as happens for the series expansion of a realistic band dispersion), we are guaranteed to have a smooth coordinate transformation with a smooth inverse $\psi : (\tilde{k}_x, \tilde{k}_y) \rightarrow(k_x, k_y)$ that maps the full dispersion to the exact form of the HOVHS in some neighborhood of the critical point. That is, with $(k_x, k_y) = \psi (\tilde{k}_x, \tilde{k_y})$ we have $E(k_x, k_y) = E(\psi(\tilde{k}_x, \tilde{k}_y)) = \tilde{k}_x^4 - \tilde{k}_y^4$ (or the equivalent form for the other HOVHS). The DOS integral then takes the form
\begin{align}
    g(\varepsilon) & \approx \int_{\mathbb{R}^{2}} \frac{d^2 \tilde{k}}{(2 \pi)^2} \, J ( \tilde{k}_x, \tilde{k}_y ) \, \delta ( \tilde{k}_x^4 - \tilde{k}_y^2 - \varepsilon ),    
\end{align}
where $J(\tilde{k}_x, \tilde{k}_y)$ is the Jacobian determinant for $\psi$ that is necessary to effect the change of variables in a multiple integral. Since $\psi$ is a local diffeomorphism, $J_0 = J(0,0)$ has to be non-zero. We can then series expand $J(\tilde{k}_x, \tilde{k}_y)$ around $(0,0)$, scale $\tilde{k}_x$ and $\tilde{k}_y$ in each integral in the resultant series appropriately to obtain the power-law DOS at leading order, that is corrected by sub-leading terms \cite{Chandrasekaran2020}
\begin{equation}
    g(\varepsilon) \approx |\varepsilon|^{-\gamma} \! \left( \! J_{0}^{\null} D_{\pm}^{\null} \! + \! \sum\limits_{n = 1}^{\infty} \sum_{m = 0}^{n} c_{mn}^{\null} |\varepsilon|^{m \alpha + (n-m) \beta} \! \right),
\end{equation}
where the leading order exponent $\gamma = 1 - \alpha - \beta$. Since $0 < \alpha, \beta < 1$ and $m \leq n$, we have $m \alpha + (n-m) \beta > 0$, which ensures that terms in the sum are subleading in the $\varepsilon \rightarrow 0$ limit. While fitting for the power law DOS arising from a realistic dispersion with HOVHS, we may have to include a few terms in this series to get an accurate description for the chosen energy range near the critical energy. Notice that the leading order terms for $\varepsilon > 0$ and $\varepsilon < 0$ respectively have the coefficients $J_0 D_+$ and $J_0 D_-$ ensuring that the universal ratio of prefactors $D_+ / D_-$ is preserved.

%


\clearpage

\begin{widetext}

\unappendix

\section*{Supplemental Material for ``Designing Topological High-Order van Hove Singularities: Twisted Bilayer Kagom\'{e}''}

\FloatBarrier

\setcounter{section}{0}
\setcounter{equation}{0}
\setcounter{figure}{0}
\setcounter{table}{0}
\setcounter{page}{1}

\renewcommand{\theequation}{S\arabic{equation}}
\renewcommand{\thesection}{S\arabic{section}}
\renewcommand{\thesubsection}{S\arabic{section}.\arabic{subsection}}
\renewcommand{\thesubsubsection}{S\arabic{section}.\arabic{subsection}.\arabic{subsubsection}}
\renewcommand{\thefigure}{S\arabic{figure}}
 \renewcommand{\thetable}{S\arabic{table}}

\section{Monolayer and Bilayer Kagom\'{e}}

\subsection{Monolayer Kagom\'{e}}

To introduce conventions and help familiarize ourselves with the Kagom\'{e} system, let us briefly recap the monolayer Kagom\'{e} system. We present a schematic of the Kagom\'{e} lattice, Brillouin zone (BZ), and (reciprocal) lattice vectors in Fig. \ref{Kagome_monolayer_schematic}. The lattice vectors and sublattice positions are taken to be
\begin{equation}
    \mathbf{a}_{1} = a \colVec{1,0}, \quad \mathbf{a}_{2} = \frac{a}{2} \colVec{1,\sqrt{3}}, \quad \boldsymbol{\delta}_{a} = \frac{a}{2\sqrt{3}} \colVec{0,1}, \quad \boldsymbol{\delta}_{b} = \frac{a}{4\sqrt{3}} \colVec{\sqrt{3},-1}, \quad \boldsymbol{\delta}_{c} = \frac{a}{4\sqrt{3}} \colVec{-\sqrt{3},-1},
\end{equation}
with the origin set as the centre of an up triangle. The corresponding reciprocal lattice vectors are then
\begin{equation}
    \mathbf{b}_{1} = \frac{2\pi}{\sqrt{3}a} \colVec{\sqrt{3},-1}, \qquad \mathbf{b}_{2} = \frac{4\pi}{\sqrt{3}a} \colVec{0,1}.
\end{equation}

\begin{figure}
    \centering
    \includegraphics[width=0.85\textwidth]{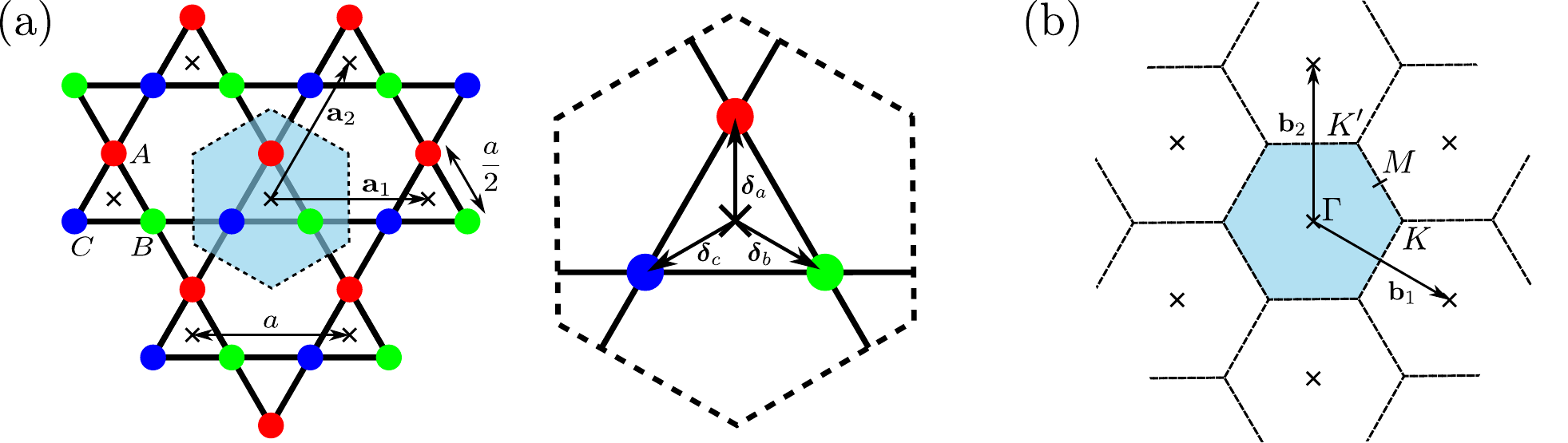}
    \caption{(a): Kagom\'{e} lattice structure with the lattice vectors, lattice sites (crosses), sublattices, and unit cell (shaded region) identified. (b): Reciprocal lattice of the Kagom\'{e} lattice with the reciprocal lattice points marked by crosses. Wigner-Seitz unit cells are marked by dotted lines and the first Brillouin zone is shaded.}
    \label{Kagome_monolayer_schematic}
\end{figure}

The tight binding Hamiltonian for the Kagom\'{e} lattice with a single orbital per site, assuming constant nearest-neighbor tunneling, is simply
\begin{equation}
\begin{split}
    H &= -t \sum_{i} \left[ (b_{i}^{\dagger} + c_{i}^{\dagger}) a_{i}^{\null} + c_{i}^{\dagger} b_{i}^{\null} \right] - t \sideset{}{'} \sum_{\angleaverage{i,j}} \left[ (b_{j}^{\dagger} + c_{j}^{\dagger}) a_{i}^{\null} + c_{j}^{\dagger} b_{i}^{\null} \right] + h.c. \\
    &= -t \sum_{i} \sum_{\eta \neq \chi} \eta_{i}^{\dagger} \chi_{i}^{\null} -t \sideset{}{'}\sum_{\angleaverage{i,j}} \sum_{\eta \neq \chi} \eta_{j}^{\dagger} \chi_{i}^{\null}, \qquad \chi,\eta \in \{a,b,c\},
    \label{Kagome_monolayer_TB_Hamiltonian}
\end{split}
\end{equation}
where $t$ is the tunneling energy, $\eta_{i}^{\dagger}$ and $\chi_{i}^{\dagger}$ are creation operators for sublattice $\eta$ and $\chi$, respectively, for the unit cell located at $\mathbf{R}_{i}$, $\eta_{i}^{\null}$ and $\chi_{i}^{\null}$ are the corresponding annihilation operators, $\angleaverage{i,j}$ denotes nearest-neighbor unit cells, and the primed sum indicates the restriction of the sum to ensure the sublattice sites are nearest-neighbors. We move to momentum space by performing a change of basis for the creation and annihilation operators,
\begin{equation}
    \eta_{i}^{\null} = \frac{1}{\sqrt{N}} \sum_{\mathbf{k}} e^{i \mathbf{k} \cdot (\mathbf{R}_{i} + \boldsymbol{\delta}_{\eta})} \eta_{\mathbf{k}}^{\null}, \qquad \chi_{i}^{\null} = \frac{1}{\sqrt{N}} \sum_{\mathbf{k}} e^{i \mathbf{k} \cdot (\mathbf{R}_{i} + \boldsymbol{\delta}_{\chi})} \chi_{\mathbf{k}}^{\null},
\end{equation}
in which $N$ is the number of unit cells in the periodic system. Substituting these expressions into eq. \ref{Kagome_monolayer_TB_Hamiltonian} yields
\begin{equation}
    H = -t \sum_{\mathbf{k}} \sum_{\eta \neq \chi} \gamma_{\eta\chi}^{\null}(\mathbf{k}) \eta_{\mathbf{p}}^{\dagger} \chi_{\mathbf{k}}^{\null},
    \label{Kagome_monolayer_TB_kspace_second_quantised_form}
\end{equation}
with $\gamma_{\eta\chi}(\mathbf{k}) = 2\cos(\mathbf{k} \cdot \boldsymbol{\delta}_{\eta\chi})$ being the structure factor connecting sublattices $\eta$ and $\chi$ and $\boldsymbol{\delta}_{\eta\chi} = \boldsymbol{\delta}_{\eta} - \boldsymbol{\delta}_{\chi}$. Defining the 3-component annihilation operator $\psi_{\mathbf{k}}^{\null} = (a_{\mathbf{k}}^{\null}, b_{\mathbf{k}}^{\null}, c_{\mathbf{k}}^{\null})^{T}$, we may write the Hamiltonian compactly as $H = \sum_{\mathbf{k}} \psi_{\mathbf{k}}^{\dagger} H_{\mathbf{k}}^{\null} \psi_{\mathbf{k}}^{\null}$, where
\begin{equation}
    H_{\mathbf{k}}^{\null} = \begin{pmatrix}
        0 & -t \gamma_{ab}^{\null} (\mathbf{k}) & -t \gamma_{ac}^{\null} (\mathbf{k}) \\
        -t \gamma_{ab}^{\null} (\mathbf{k}) & 0 & -t \gamma_{bc}^{\null} (\mathbf{k}) \\
        -t \gamma_{ac}^{\null} (\mathbf{k}) & -t \gamma_{bc}^{\null} (\mathbf{k}) & 0
    \end{pmatrix},
    \label{Kagome_monolayer_continuum_Hamiltonian}
\end{equation}
Finally, diagonlaising this Hamiltonian yields
\begin{equation}
    E_{0} = 2t, \qquad E_{\pm}(\mathbf{k}) = -t \left[ 1 \pm \sqrt{3 + 2\cos(2\mathbf{k}\cdot\boldsymbol{\delta}_{ab}) + 2\cos(2\mathbf{k}\cdot\boldsymbol{\delta}_{ac}) + 2\cos(2\mathbf{k}\cdot\boldsymbol{\delta}_{bc})} \right].
\end{equation}

\subsection{Dimerisation}

To introduce dimerisation into the Kagom\'{e} lattice, we include an additional tunneling that enhances hopping along up triangles and reduces hopping via down triangles, see Fig. 1c of the main text. We achieve this by adding
\begin{equation}
\begin{split}
    H_{1} &= -t_{1} \sum_{i} \left[ (b_{i}^{\dagger} + c_{i}^{\dagger}) a_{i}^{\null} + c_{i}^{\dagger} b_{i}^{\null} \right] + t_{1} \sideset{}{'}\sum_{\angleaverage{i,j}} \left[ (b_{j}^{\dagger} + c_{j}^{\dagger}) a_{i}^{\null} + c_{j}^{\dagger} b_{i}^{\null} \right] + h.c. \\
    &= -t_{1} \sum_{i} \sum_{\eta \neq \chi} \eta_{i}^{\dagger} \chi_{i}^{\null} + t_{1} \sideset{}{'}\sum_{\angleaverage{i,j}} \sum_{\eta \neq \chi} \eta_{j}^{\dagger} \chi_{i}^{\null}.
    \label{Kagome_monolayer_dimerising_TB_Hamiltonian}
\end{split}
\end{equation}
to the bare monolayer Hamiltonian, $H$. Written in momentum space, this contributes
\begin{equation}
    H_{1,\mathbf{k}}^{\null} = i t_{1}^{\null} \begin{pmatrix}
        0 & -\tilde{\gamma}_{ab}^{\null} (\mathbf{k}) & -\tilde{\gamma}_{ac}^{\null} (\mathbf{k}) \\
        \tilde{\gamma}_{ab}^{\null} (\mathbf{k}) & 0 & -\tilde{\gamma}_{bc}^{\null} (\mathbf{k}) \\
        \tilde{\gamma}_{ac}^{\null} (\mathbf{k}) & \tilde{\gamma}_{bc}^{\null} (\mathbf{k}) & 0
    \end{pmatrix},
\end{equation}
where $\tilde{\gamma}_{\eta\chi}^{\null} (\mathbf{k}) = 2 \sin(\mathbf{k} \cdot \boldsymbol{\delta}_{\eta\chi})$. The bands for the full Hamiltonian, $H + H_{1}$, are then just
\begin{equation}
    E_{0} = 2t, \qquad E_{\pm}(\mathbf{k}) = -t \pm \sqrt{3 t^{2} + 6 t_{1}^{2} + 2(t^{2} - t_{1}^{2}) ( \cos(2\mathbf{k}\cdot\boldsymbol{\delta}_{ab}) + \cos(2\mathbf{k}\cdot\boldsymbol{\delta}_{ac}) +  \cos(2\mathbf{k}\cdot\boldsymbol{\delta}_{bc}) )}.
\end{equation}
Consequently, at the $K$ point the Dirac cone becomes gapped, $E_{\pm}(\mathbf{K}) = -t \pm t_{1}$, similar to the inclusion of a mass term in the low-energy description of monolayer graphene.

\subsection{Haldane-type Hopping}

Choosing to instead include a complex next-nearest-neighbor tunneling mechanism \cite{Wang2025}, we define anti-clockwise tunneling to possess a phase factor of $e^{i\phi}$ and clockwise tunneling to possess $e^{-i\phi}$, see Fig. 1d of the main text. The Hamiltonian describing this process is
\begin{equation}
\begin{split}
    H_{2} = -t_{2} \sideset{}{''}\sum_{\angleaverage{i,j}} \left[ e^{i\phi} (a_{i}^{\dagger}b_{j}^{\null} + b_{i}^{\dagger}c_{j}^{\null} + c_{i}^{\dagger}a_{j}^{\null}) + h.c. \right] = -t_{2} \sideset{}{''}\sum_{\angleaverage{i,j}} \sum_{\eta \neq \chi} e^{i\mathcal{S}_{\eta\chi}\phi} \eta_{i}^{\dagger} \chi_{j}^{\null},
    \label{Kagome_monolayer_Haldane_TB_Hamiltonian}
\end{split}
\end{equation}
where the double primed sum denotes restriction to ensure the sublattices are next-nearest-neighbor and $\mathcal{S}_{\eta\chi} = \pm 1$ determines the sign of the complex tunneling phase. By inspecting eq. \ref{Kagome_monolayer_Haldane_TB_Hamiltonian}, we see a cyclic permutation of the sublattices between the creation and annihilation operators that possess the same phase factor. We may therefore map the sublattice labels $\{A,B,C\} \rightarrow \{1,2,3\}$ to write the phase sign compactly in terms of the rank-3 Levi-Civita tensor,
\begin{equation}
    \mathcal{S}_{\eta\chi} = \sum_{\alpha = 1}^{3} \varepsilon_{\eta\chi\alpha}.
\end{equation}
Finally, moving to momentum space, we arrive at
\begin{equation}
    H_{2,\mathbf{k}}^{\null} = -t_{2}^{\null} \begin{pmatrix}
        0 & e^{i\phi} \bar{\gamma}_{ab}^{\null} (\mathbf{k}) & e^{-i\phi} \bar{\gamma}_{ac}^{\null} (\mathbf{k}) \\
        e^{-i\phi} \bar{\gamma}_{ab}^{\null} (\mathbf{k}) & 0 & e^{i\phi} \bar{\gamma}_{bc}^{\null} (\mathbf{k}) \\
        e^{i\phi} \bar{\gamma}_{ac}^{\null} (\mathbf{k}) & e^{-i\phi} \bar{\gamma}_{bc}^{\null} (\mathbf{k}) & 0
    \end{pmatrix},
\end{equation}
where $\bar{\gamma}_{ij}^{\null}(\mathbf{k})$ is the bare monolayer structure factor shifted by an appropriate lattice vector,
\begin{equation}
    \bar{\gamma}_{ab}^{\null} (\mathbf{k}) = 2\cos(\mathbf{k} \cdot (\boldsymbol{\delta}_{ab}^{\null}+\mathbf{a}_{1}^{\null})), \quad \bar{\gamma}_{ab}^{\null} (\mathbf{k}) = 2\cos(\mathbf{k} \cdot (\boldsymbol{\delta}_{ac}^{\null}-\mathbf{a}_{1}^{\null})), \quad \bar{\gamma}_{bc}^{\null} (\mathbf{k}) = 2\cos(\mathbf{k} \cdot (\boldsymbol{\delta}_{bc}^{\null}-\mathbf{a}_{2}^{\null})).
\end{equation}

\subsection{Untwisted Bilayer Hamiltonians}

\subsubsection{$AA$ Kagom\'{e} Bilayer}

For this choice of stacking, we introduce the interlayer tunneling strength $t_{\perp}$ to characterise the energy scale associated to the coupling between the layers. The effective tight binding Hamiltonian can then be written as
\begin{equation}
    H_{AA} = H_{1} + H_{2} + H_{T,AA}, \qquad H_{T,AA} = -t_{\perp} \sum_{i} \sum_{\eta} \left[ \eta_{i,1}^{\dagger} \eta_{i,2}^{\null} + \eta_{i,2}^{\dagger} \eta_{i,1}^{\null} \right],
\end{equation}
where $H_{1,2}$ are the isolated Kagom\'{e} Hamiltonians for the first and second layers, respectively, $H_{T,AA}$ is the interlayer tunneling Hamiltonian, and $\eta_{i,\alpha}^{\dagger}$ ($\eta_{i,\alpha}^{\null}$) creates (annihilates) an electron on sublattice $\eta$ of the $\alpha$\textsuperscript{th} layer in the $i$\textsuperscript{th} unit cell. Moving to momentum space we find in the basis $\psi_{\mathbf{k}}^{\null} = (a_{1\mathbf{k}}^{\null}, b_{1\mathbf{k}}^{\null}, c_{1\mathbf{k}}^{\null}, a_{2\mathbf{k}}^{\null}, b_{2\mathbf{k}}^{\null}, c_{2\mathbf{k}}^{\null})^{\text{T}}$,
\begin{equation}
    H_{\mathbf{k}}^{(AA)} = \begin{pmatrix}
        H_{\mathbf{k}}^{(1)} & H_{T}^{(AA)} \\
        H_{T}^{(AA)} & H_{\mathbf{k}}^{(2)}
    \end{pmatrix}, \qquad H_{T}^{(AA)} = \begin{pmatrix}
        -t_{\perp} & 0 & 0 \\
        0 & -t_{\perp} & 0 \\
        0 & 0 & -t_{\perp}
    \end{pmatrix}.
\end{equation}
Due to the momentum being 2D, the interlayer separation distance, $d_{\perp}$, does not appear explicitly in this Hamiltonian. The bands for the $AA$ bilayer system are then found to simply be those of the monolayer system shifted by $\pm t_{\perp}$,
\begin{equation}
    E_{0,\pm} = 2t \pm t_{\perp}, \qquad E_{\mu\nu} = -\mu t_{\perp} - t \left[ 1 + \nu \sqrt{3 + 2\cos(2\mathbf{k}\cdot\boldsymbol{\delta}_{ab}) + 2\cos(2\mathbf{k}\cdot\boldsymbol{\delta}_{ac}) + 2\cos(2\mathbf{k}\cdot\boldsymbol{\delta}_{bc})} \right],
\end{equation}
where $\mu,\nu = \pm 1$ label the different dispersive bands, which are shown in Fig. 1b of the main text.

\subsubsection{$AB$ Kagom\'{e} Bilayer}

With the same notation as before, the only change in this case is that tunneling may only occur between the $A$ and $B$ sublattices of the two layers, meaning there will be no matrix elements connecting the $C$ sites between layeres. The Hamiltonian for this system is then simply
\begin{equation}
    H_{AB} = H_{1} + H_{2} + H_{T,AB}, \qquad H_{T,AB} = -t_{\perp} \sum_{i} \left[ a_{i,1}^{\dagger} b_{i,2}^{\null} + b_{i,1}^{\dagger} a_{i,2}^{\null} + h.c. \right].
\end{equation}
Moving to momentum space, we find
\begin{equation}
    H_{\mathbf{k}}^{(AB)} = \begin{pmatrix}
        H_{\mathbf{k}}^{(1)} & H_{T}^{(AB)} \\
        H_{T}^{(AB)} & H_{\mathbf{k}}^{(2)}
    \end{pmatrix}, \qquad H_{T}^{(AB)} = \begin{pmatrix}
        0 & -t_{\perp} & 0 \\
        -t_{\perp} & 0 & 0 \\
        0 & 0 & 0
    \end{pmatrix},
\end{equation}
The eigenvalues of this Hamiltonian cannot be found analytically and so we resort to numerical diagonalisation to obtain the band structure shown in Fig. 1b of the main text.

\subsubsection{Interlocked Kagom\'{e} Bilayer}

In this scenario where no sublattice site of one layer is located directly above any sublattice of the other layer, we instead focus on the partial overlap between sublattices of the two layers. For example, the $A$ sites of one layer will have a reduced tunneling to the $B$ and $C$ sites of the other layer compared to the $AA$ and $AB$ stacking choices. Let us denote this reduced tunneling energy by $\tilde{t}_{\perp}$. We may then write the tight binding Hamiltonian for this system as
\begin{equation}
    H_{\text{Int}} = H_{1} + H_{2} + H_{T,\text{Int}}, \qquad H_{T,\text{Int}} = -\tilde{t}_{\perp} \sum_{i} \sum_{l} \sum_{\eta \neq \chi} \eta_{i,\bar{l}}^{\dagger} \chi_{i,l}^{\null},
\end{equation}
where the sum over $l$ is the sum over layers and $\bar{l}$ refers to the opposite choice of $l$ (i.e. if $l = 1$ then $\bar{l} = 2$ and vice versa). Unlike the $AA$ and $AB$ bilayers, we find that moving to momentum space generates an additional momentum dependent matrix elements of the continuum Hamiltonian due to the offset of the sublattices of the two layers of the form $\Phi_{\eta\chi}(\mathbf{k}) = e^{-i\mathbf{k} \cdot \boldsymbol{\delta}_{\eta_{1}\chi_{2}}}.$ The continuum Hamiltonian for this system can then be written compactly as
\begin{equation}
    H_{\mathbf{k}}^{(\text{Int})} = \begin{pmatrix}
        H_{\mathbf{k}}^{(1)} & H_{T,\mathbf{k}}^{(\text{Int})} \\
        H_{T,\mathbf{k}}^{(\text{Int})\dagger} & H_{\mathbf{k}}^{(2)}
    \end{pmatrix}, \qquad H_{T,\mathbf{k}}^{(\text{Int})} = \begin{pmatrix}
        0 & -\tilde{t}_{\perp} \Phi_{ab}(\mathbf{k}) & -\tilde{t}_{\perp} \Phi_{ac}(\mathbf{k})  \\
        -\tilde{t}_{\perp} \Phi_{ba}(\mathbf{k}) & 0 & -\tilde{t}_{\perp} \Phi_{bc}(\mathbf{k}) \\
        -\tilde{t}_{\perp} \Phi_{ca}(\mathbf{k}) & -\tilde{t}_{\perp} \Phi_{cb}(\mathbf{k}) & 0
    \end{pmatrix}.
\end{equation}
As in the $AB$ stacking case, we find ourselves again unable to obtain analytic expressions for the band structure of the interlocked bilayer and obtain the band structure shown in Fig. 1b of the main text numerically.

\section{Comparison of TBK Structures}

In the main text we saw that the band structure of TBK was sensitive to the choice of stacking order primarily around the $K_{\text{M}}$ and $M_{\text{M}}$ points outside of the near-flat band region in the absence of dimerisation and Haldane hopping. When either of these effects are introduced into the Hamiltonian we find that the various stacking choices start to exhibit clearer differences in their band structures, as shown in Fig. \ref{Band_comparison}. The largest changes between the stacking orders are seen in the lower 14 bands (i.e. those below the MDP energy), especially about the $M_{\text{M}}$ point, though notable differences can still be seen along the high-symmetry path between $\Gamma_{\text{M}}$ and $K_{\text{M}}$ and at the MBZ corners.

\begin{figure}
    \centering
    \includegraphics[width=\textwidth]{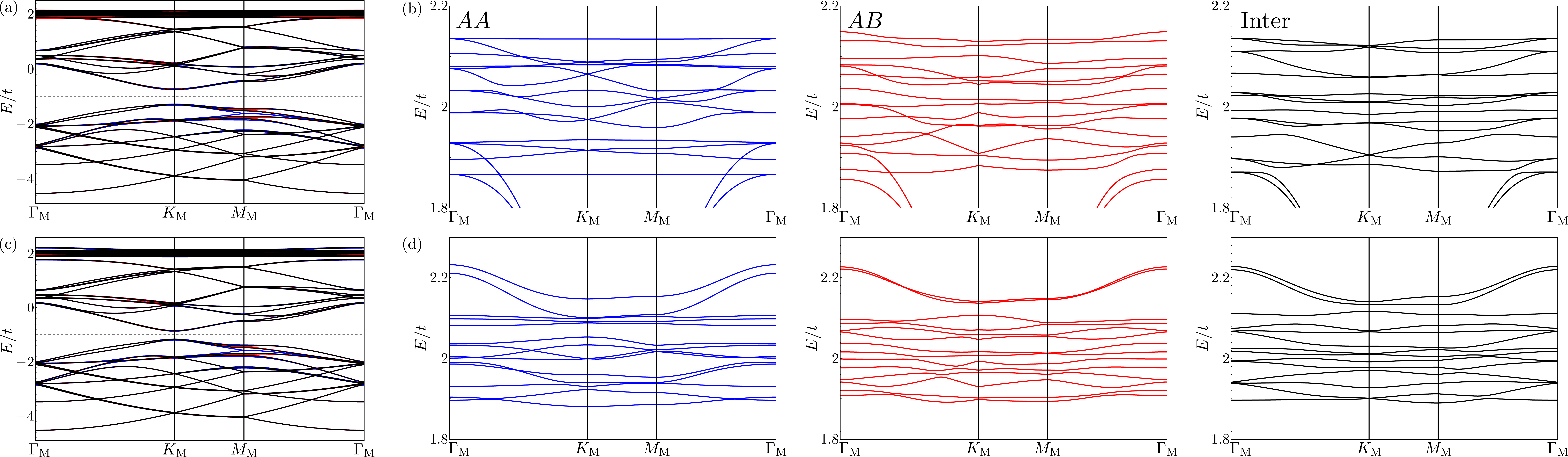}
    \caption{Comparison of the band structures for the three high-symmetry stacking orders in the presence of dimerisation and Haldane hopping. (a): Complete band structure of all three stackings with $t_{d} = 0.1t$. (b): Near-flat band region of panel (a) for each stacking order. (c): Completed band structure of all three stackings with $t_{\text{H}} = 0.05 t$ and $\phi = \pi/2$. (d): Near-flat band region of panel (c) for each stacking order. In all cases above we have taken $\theta_{c} = 38.2^{\circ}$, $a = 0.5338$ nm, $d_{\perp} = 0.6596$ nm, $t_{\text{H}} = 0$, $\phi = 0$, $t_{\perp} = 0.3 t$, $\gamma = 20$ \cite{Ye2018,Lima2019}. The grey dashed line indicates the MDP energy.}
    \label{Band_comparison}
\end{figure}

When dimerisation is introduced in all cases (Figs. \figRef{Band_comparison}{a} and \figRef{Band_comparison}{b}), we find that the relatively flat nature of the bands is preserved with the bands becoming slightly more dispersive and the width of the near-flat band region is increased. In particular, we see what appear to be two flat energy surfaces appearing in the twisted $AA$ bilayer, reminiscent of the untiwsted case. However, upon closer inspection, we find this is not actually the case. For the uppermost band (band 42), we find it exhibits variations in energy on the scale of $\sim 10^{-3} t$. In contrast the lower apparently flat surface formed of two bands near $E = 1.86 t$ exhibits an avoided crossing and not a degeneracy between the two bands the surface lies in. Moreover, we find a similarly weak momentum dependence yielding variations in energy on the scale of $\sim 10^{-4} t$. If we instead use an imaginary next-nearest-neighbor tunneling, as in Figs. \figRef{Band_comparison}{c} and \figRef{Band_comparison}{d}, we find that bands 41 and 42 become significantly more dispersive for all choices of stacking order. However, the choice of stacking can be seen to be crucial in whether or not an insulating gap opens and in the size of this gap. This suggests that the stacking order will play a central role in transport measurements of Hall conductance due to topologically protected edge states.

\begin{figure}[t]
    \centering
    \includegraphics[width=0.85\textwidth]{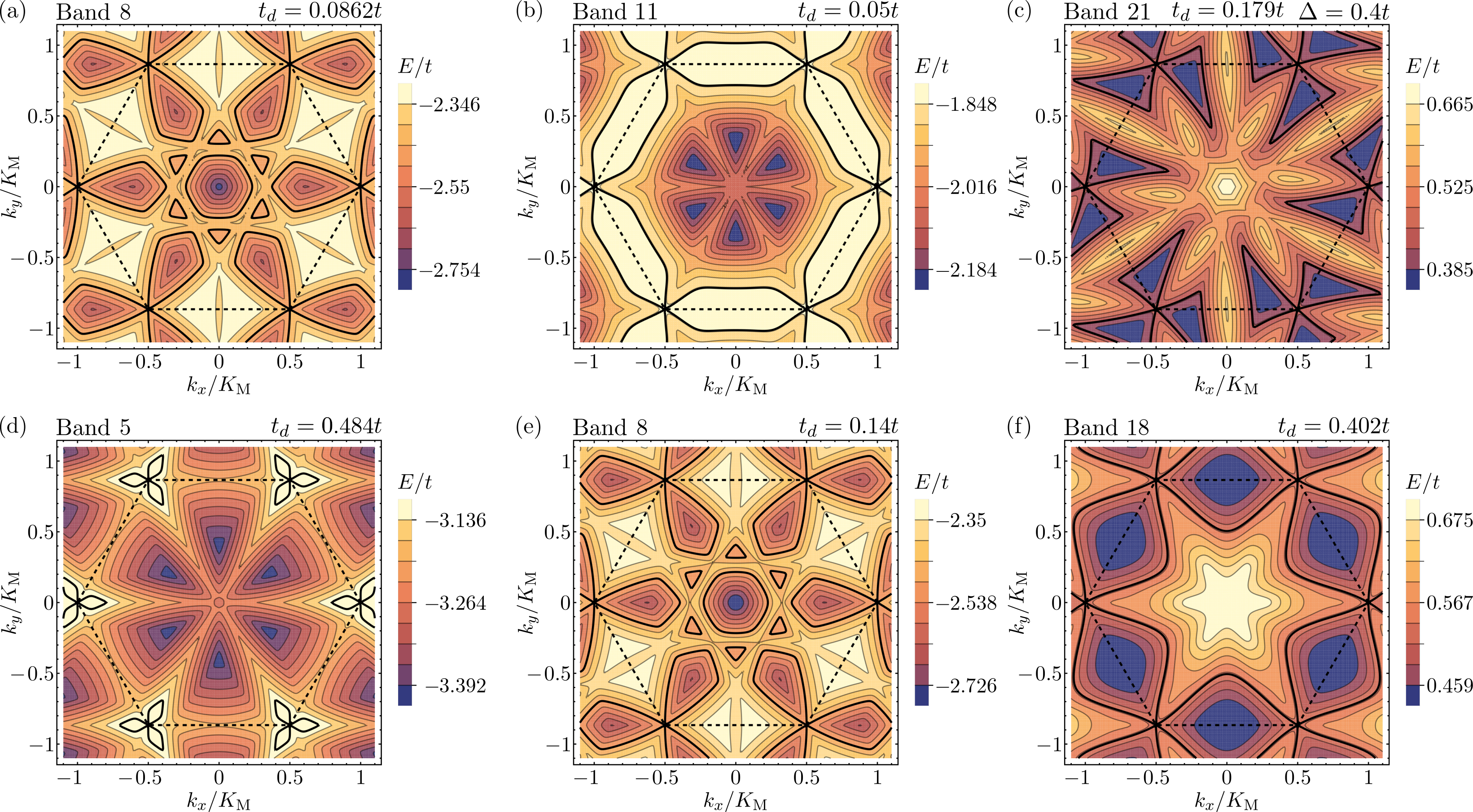}
    \caption{(a)-(c): Monkey saddle singularities arising in $AA$ TBK. (d)-(f): Monkey saddle singularities appearing in interlocked TBK. The band considered and choice of dimerisation is stated in each panel. Only in (c) is the interlayer potential taken to be non-zero, thus breaking the effective $k_{y} \rightarrow -k_{y}$ symmetry. The bold black lines denote the energy contour of the critical point, while the dashed line indicates the MBZ boundary. We take $\theta_{c} = 38.2^{\circ}$, $a = 0.5338$ nm, $d_{\perp} = 0.6596$ nm, $t_{\text{H}} = 0$, $\phi = 0$, $t_{\perp} = 0.3 t$, $\gamma = 20$ \cite{Ye2018,Lima2019} for the system considered here.}
    \label{MS_dim_examples}
\end{figure}

\begin{figure}[H]
    \centering
    \includegraphics[width=0.85\textwidth]{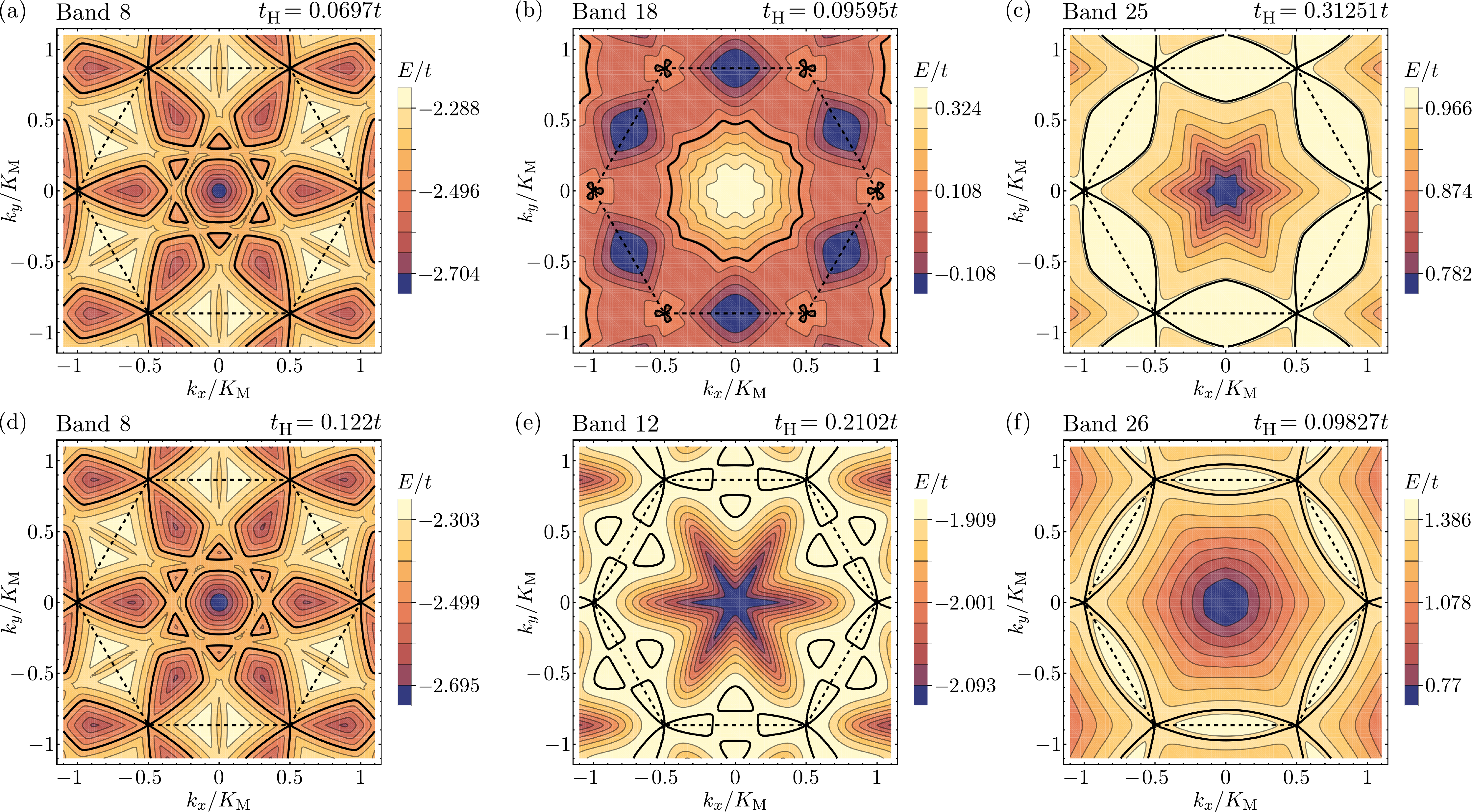}
    \caption{(a)-(c): Monkey saddle singularities arising in $AA$ TBK. (d)-(f): Monkey saddle singularities appearing in interlocked TBK. The band considered and choice of Haldane hopping is stated in each panel. The bold black lines denote the energy contour of the critical point, while the dashed line indicates the MBZ boundary. We take $\theta_{c} = 38.2^{\circ}$, $a = 0.5338$ nm, $d_{\perp} = 0.6596$ nm, $t_{d} = 0$, $\phi = \pi/2$, $t_{\perp} = 0.3 t$, $\gamma = 20$ \cite{Ye2018,Lima2019} for the system considered here.}
    \label{MS_topo_examples}
\end{figure}

\section{Monkey Saddle Singularities}

\subsection{Dimerisation}

We present further examples of monkey saddle singularities for both the $AA$ and interlocked TBK systems to highlight the wide range of parameter choices that may be taken in engineering monkey saddles. Fig. \ref{MS_dim_examples} contains monkey saddles created through dimerisation of the Kagom\'{e} lattice. Here we see that the choice of stacking order can significantly change the dimerisation required to obtain a singularity around the $K_{\text{M}}$ point, as is the case for band 8 (Fig. \figRef{MS_dim_examples}{a} and Fig. \figRef{MS_dim_examples}{e}). Moreover, we find that an interlayer potential can act as an additional tuning knob to create monkey saddles at the MBZ corners, see Fig. \figRef{MS_dim_examples}{b}. We also observe that both localized and delocalized singularities manifest in other bands, highlighting that both can be found with relative ease. A list of the monkey saddles we have found TBK is given in Table \ref{Monkey_saddle_table} alongside the dimerisations and interlayer potentials required to engineer them.

\subsection{Topological tunneling}

As discussed in the main text, we may instead engineer the TBK band structure through the introduction of a complex next-nearest-neighbor hopping to create monkey saddle singularities. Here we present further examples of monkey saddle singularities that can be acquired in both $AA$ and interlocked TBK through a purely imaginary Haldane tunneling term. Fig. \ref{MS_topo_examples} illustrates the range of values that may be taken for $t_{\text{H}}$ and how monkey saddles can be acquired in several bands outside of the near-flat band region. We provide a list of the monkey saddle singularities we have found in TBK due to Haldane hopping in Table \ref{Monkey_saddle_table}.

\begin{table}
\centering
\caption{Summary of the approximate parameters yielding monkey saddle singularities in $AA$ and interlocked TBK when either dimerisation of Haldane hopping is present. In all cases we take $\theta_{c} = 38.2^{\circ}$, $a = 0.5338$ nm, $d_{\perp} = 0.6596$ nm, $t_{\perp} = 0.3 t$, $\gamma = 20$ \cite{Ye2018,Lima2019}. The parameters listed here should be read as two sets: unbracketed values belong to $AA$ TBK while bracketed values belong to interlocked TBK. The asterisk indicates the only case where an interlayer potential is present with $\Delta = 0.4t$.}
\label{Monkey_saddle_table}
\begin{tabular}{ c c | c c || c c | c c }
    \hline
        \multicolumn{4}{c ||}{Dimerisation} & \multicolumn{4}{c}{Topological} \\
        Band & $t_{d}/t$ & Band & $t_{d}/t$ & Band & $t_{\text{H}}/t$ & Band & $t_{\text{H}}/t$ \\ 
    \hline\hline
        2 & 0.68 (0.7) & 18 & 0.29 (0.402) & 2 & 0.621 (0.66) & 18 & 0.09595 \\
        5 & 0.5 (0.484) & 21 & $0.179^{*}$ & 5 & 0.4859 (0.5172) & 20 & 0.18147 (0.26589) \\
        8 & 0.0862 (0.14) & 25 & 0.51 (0.313) & 8 & 0.0697 (0.122) & 21 & (0.0418) \\
        11 & 0.05 (0.14) & 26 & 0.44533 (0.525) & 11 & 0.04498 (0.1188) & 25 & 0.31251 (0.28619) \\
        12 & 0.59 (0.3195) & -- & -- & 12 & (0.2102) & 26 & 0.470977 (0.09825) \\
    \hline
\end{tabular}
\end{table}

\section{Sixth-Order Singularities}

Given the extremely sensitive nature of sixth-order VHSs, it is difficult to engineer the TBK band structure to host them with only a couple of tuning parameters. We present examples of effective sixth-order singularities appearing in band 22 for both $AA$ and interlocked TBK in Fig. \ref{6thOrder_HOVHSs} using a single parameter. We find there to exists a small minimum for the interlocked stacking at $\Gamma_{\text{M}}$ surrounded by set of of six closely packed second-order VHSs. The depth of this minimum relative to the saddle point energies of these surrounding singularities is $\sim 10^{-6} t$, suggesting that we should expect divergences in the DOS characteristic of a sixth-order singularity. However, for the $AA$ configuration, we are unable to observe such a set of second-orders singularities as clearly and obtain a vanishingly small value for the second-order coefficient, $c_{0}^{(2)}$, in the expansion of the energy about the $\Gamma_{\text{M}}$ point. Given the lack of a clear extremum in our calculations, see Fig. \figRef{6thOrder_HOVHSs}{a}, this suggests the system is tuned very close to an extended extremum \cite{Chandrasekaran2025arxiv}, where the extremal value is not isolated to a single point but instead a set of connected points extending from the extremal point to infinity.

\begin{figure}[H]
    \centering
    \includegraphics[width=0.7\textwidth]{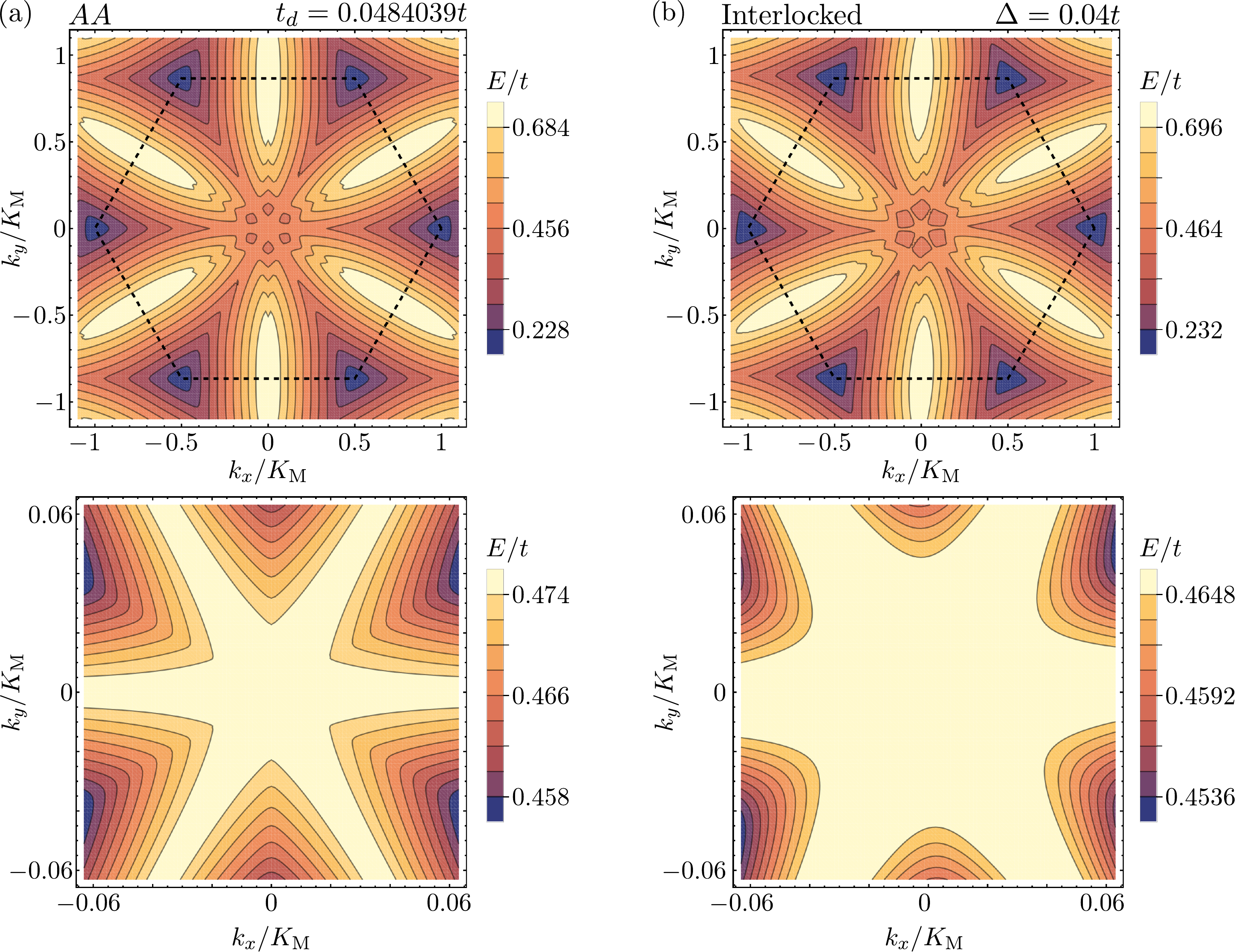}
    \caption{Examples of effective sixth-order HOVHSs in in band 22 for $AA$ (a) and interlocked (b) TBK. We take $\theta_{c} = 38.2^{\circ}$, $a = 0.5338$ nm, $d_{\perp} = 0.6596$ nm, $t_{\text{H}} = 0$, $\phi = 0$, $t_{\perp} = 0.3 t$, $\gamma = 20$ \cite{Ye2018,Lima2019}.}
    \label{6thOrder_HOVHSs}
\end{figure}

\begin{figure}[H]
    \centering
    \includegraphics[width=0.85\textwidth]{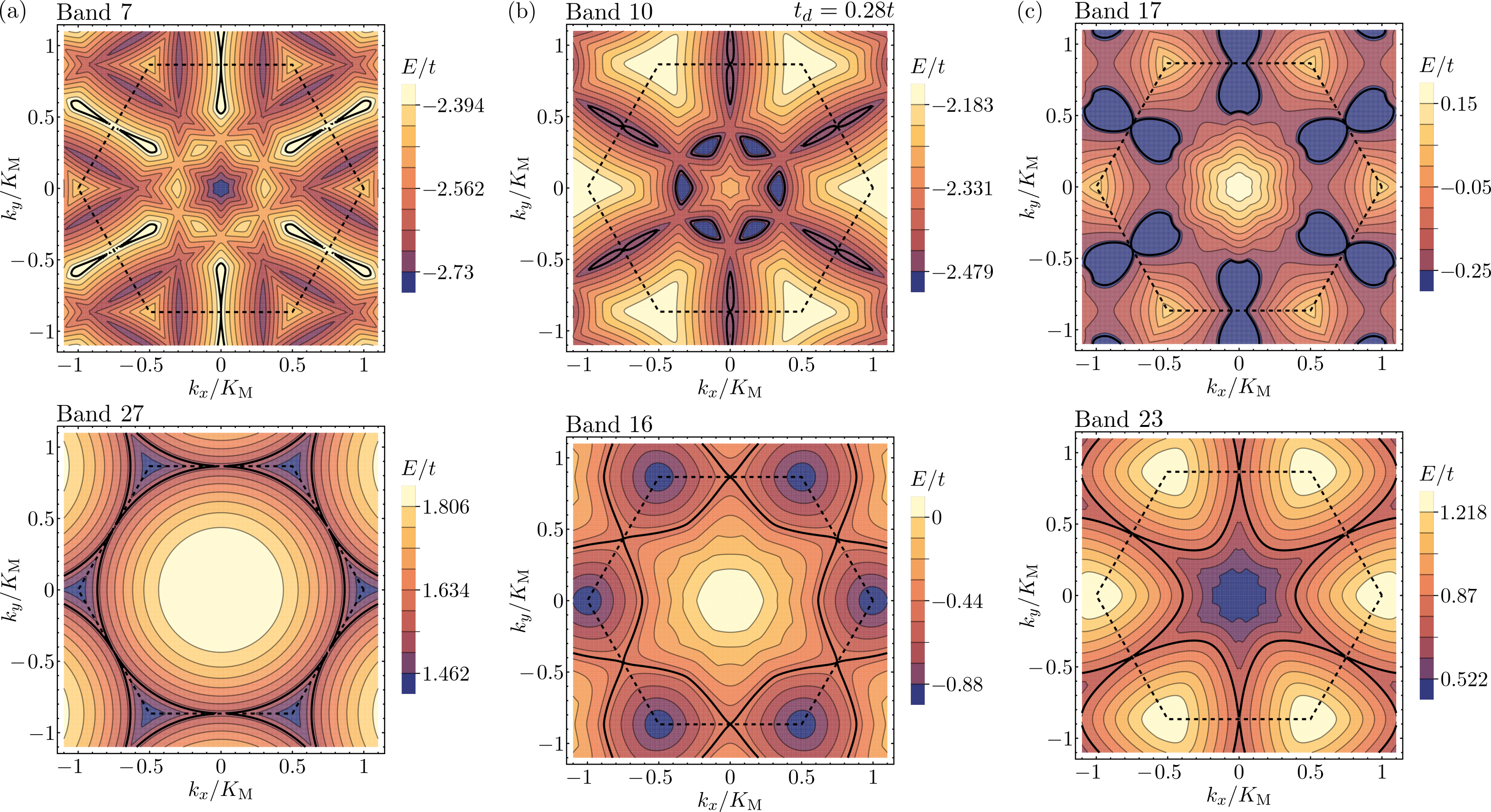}
    \caption{Examples of second-order singularities appearing at the $M_{\text{M}}$ point for $AA$ (a), interlocked (b), and $AB$ (c) stacked bilayers. Only for band 10 of interlocked TBK is a non-zero dimerisation used. The bold black lines denote the energy contour of the critical point, while the dashed line indicates the MBZ boundary. We take $\theta_{c} = 38.2^{\circ}$, $a = 0.5338$ nm, $d_{\perp} = 0.6596$ nm, $t_{\text{H}} = 0$, $\phi = 0$, $t_{\perp} = 0.3 t$, $\gamma = 20$ \cite{Ye2018,Lima2019} for the system considered here.}
    \label{2ndOrder_examples}
\end{figure}

\begin{figure}[t]
    \centering
    \includegraphics[width=0.85\textwidth]{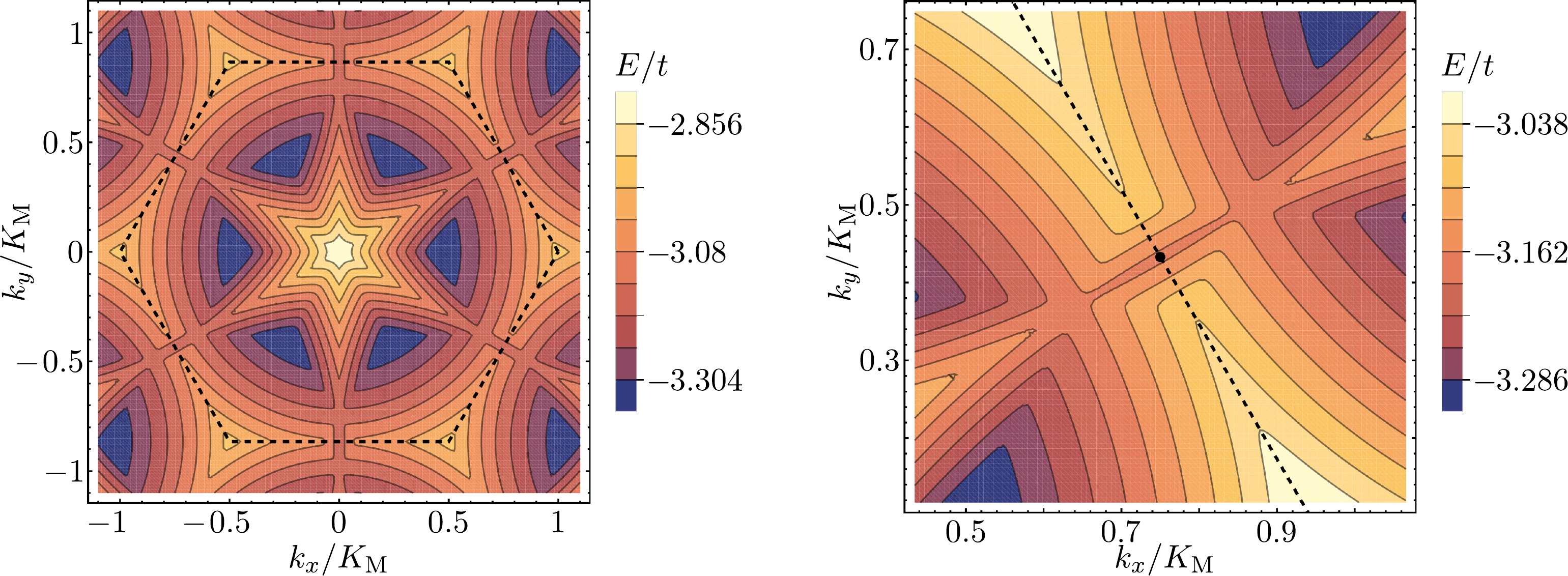}
    \caption{Potential fourth-order HOVHS about the $M_{\text{M}}$ point in band 4 of $AA$ TBK with no dimerisation. Similar features appear in band 4 of both interlocked and $AB$ as well. We take $\theta_{c} = 38.2^{\circ}$, $a = 0.5338$ nm, $d_{\perp} = 0.6596$ nm, $t_{\text{H}} = 0$, $\phi = 0$, $t_{\perp} = 0.3 t$, $\gamma = 20$ \cite{Ye2018,Lima2019}.}
    \label{4thOrder_HOVHS}
\end{figure}

\section{Second-Order Singularities}

For completeness we present examples of the various two-fold symmetric HOVHSs in Fig. \ref{2ndOrder_examples} for all three stacking choices. We see that TBK naturally hosts many such singularities without any need to tune the Hamiltonian, exhibiting both localized and delocalized singularities. Nonetheless, it is possible to engineer these second-order singularities through dimerisation, as illustrated by band 10 for interlocked TBK in Fig. \figRef{2ndOrder_examples}{b}. We can therefore expect twisted breathing Kagom\'{e} bilayer to host unique HOVHSs with 2-fold rotational symmetry that would not appear in the regular twisted bilayer. Interestingly, we find that band 4 for all stackings without dimerisation might be tunable to host a fourth-order singularity. Specifically, we find that it hosts a small minimum at the $M_{\text{M}}$ point rather than a two-fold symmetric singularity. However, two second-order singularities lie nearby to both the $M_{\text{M}}$ point and each other, see Fig. \ref{4thOrder_HOVHS}. It may therefore be possible to engineer a fourth-order singularity via other tuning parameters not considered here.

\begin{figure}[t]
    \centering
    \includegraphics[width=0.85\textwidth]{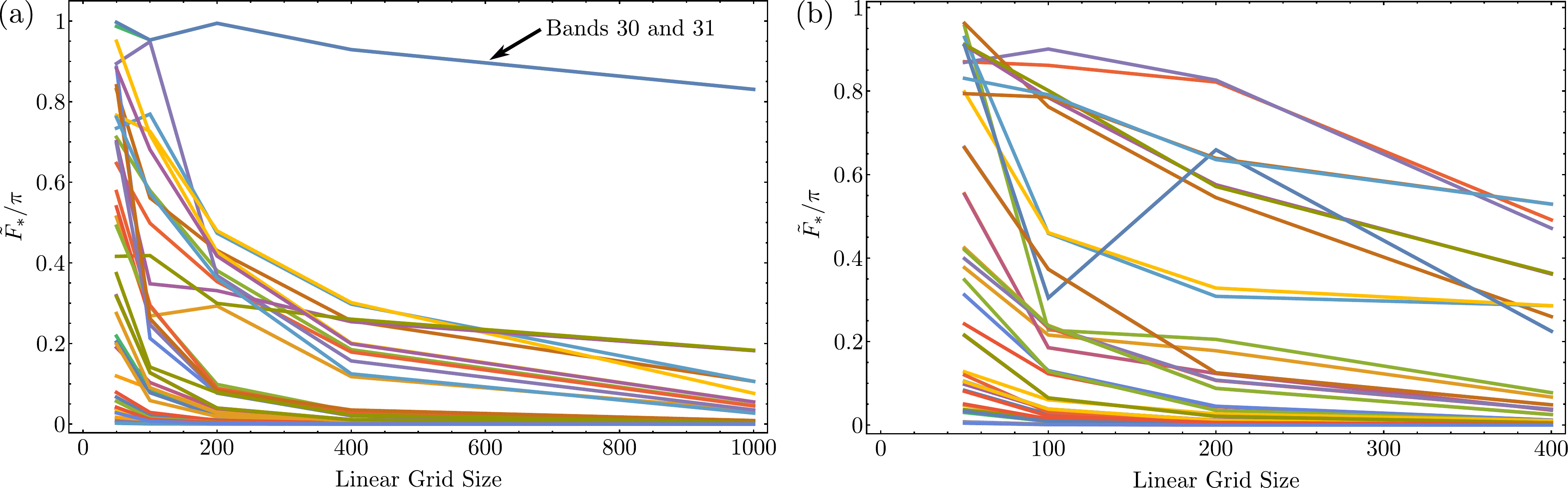}
    \caption{Variation of the maximum absolute value of the lattice field strength for the 42 bands of the $AA$ (a) and interlocked (b) TBK systems as the grid size is varied. Here we consider $\theta = 38.2^{\circ}$, $a = 0.5338$ nm, $d_{\perp} = 0.6596$ nm, $\phi = \pi/2$, $t_{\perp} = 0.3 t$, and $\gamma = 20$ \cite{Ye2018,Lima2019} with $t_{\text{H}} = 0.18147 t$ (a) and $t_{\text{H}} = 0.26589 t$ (b). Bands 30 and 31 are annotated specifically as they are what require larger grid sizes to obtain accurate calculation of the Chern numbers.}
    \label{Chern_convergence}
\end{figure}

\section{Numerical Calculation of Chern Numbers}

In calculating the Chern numbers for the 42 bands presented in Fig. 4 of the main text, we found discretising the MBZ into a $200 \times 200$ grid to be sufficient to obtain accurate results when using the method of Fukui et al. \cite{Fukui2005}. We confirmed this by performing the same calculation using both a $400 \times 400$ grid and $1000 \times 1000$ grid. We found it necessary to consider such large grid sizes due to bands 30 and 31 having an extremely small avoided crossing $\sim 10^{-6} t$. This resulted in a lattice field strength with a maximum magnitude of $\tilde{F}_{*} \sim 0.99 \pi$ for bands 30 and 31 when using smaller grid sizes, indicating the possible loss of information in the plaquettes associated to this field strength when calculating the Chern number numerically. We show the variation of $\tilde{F}_{*}$ for all 42 bands as the grid size is changed in Fig. \figRef{Chern_convergence}{a}. Here, we see its value for bands 30 and 31 reduces from the $200 \times 200$ grid onwards, reaching a value of $\tilde{F}_{*} \sim 0.83 \pi$ for the $1000 \times 1000$ grid. The Chern numbers for all 42 bands did not change between the $200 \times 200$, $400 \times 400$, and $1000 \times 1000$ grids, indicating that the Chern numbers acquired from the $200 \times 200$ had converged. Moreover, the sum of the Chern numbers vanished for the $200 \times 200$ grid and above, while their sum failed to vanish for the $100 \times 100$ and $50 \times 50$ grids also considered in Fig. \figRef{Chern_convergence}{a}. For comparison, when calculating the Chern numbers for interlocked TBK in Fig. \figRef{Chern_convergence}{b}, we found that a $100 \times 100$ grid was sufficient to capture the topological nature of the bands, as shown by the variation of $\tilde{F}_{*}$ with grid size in Fig. \figRef{Chern_convergence}{b}. We attribute this to the the use of a larger complex tunneling energy, resulting in larger gaps between the bands and avoided high concentrations of Berry curvature.

\end{widetext}

\end{document}